\documentclass[twocolumn,twocolappendix]{aastex631}
\usepackage[utf8]{inputenc}
\usepackage{natbib}
\usepackage{graphicx}
\usepackage{xcolor}
\usepackage{multirow}

\received{}
\revised{}
\accepted{}
\submitjournal{AAS Journals}

\shorttitle{Deep Dive into WASP-96\,{\rm b}}
\shortauthors{Radica et al.}

\begin{document}

\title{Super-Solar Metallicity and Tentative Evidence for Photochemistry on WASP-96\,b from JWST and Ground-Based VLT Transmission Spectroscopy}

\correspondingauthor{Michael Radica}
\email{radicamc@uchicago.edu}

\author[0000-0002-3328-1203]{Michael Radica}
\altaffiliation{NSERC Postdoctoral Fellow}
\affiliation{Department of Astronomy \& Astrophysics, University of Chicago, 5640 South Ellis Avenue, Chicago, IL 60637, USA}
\affiliation{Institut Trottier de Recherche sur les Exoplanètes, 1375 Avenue Thérèse-Lavoie-Roux, Montréal, QC, H2V 0B3, Canada}

\author[0000-0003-4844-9838]{Jake Taylor}
\affiliation{Department of Physics, University of Oxford, Parks Rd, Oxford OX1 3PU, UK}

\author[0000-0003-4459-9054]{Yoav Rotman}
\affiliation{School of Earth and Space Exploration, Arizona State University, Tempe, AZ, USA}

\author[0000-0002-0769-9614]{Jasmina Blecic}
\affiliation{Department of Physics, New York University Abu Dhabi, Abu Dhabi, UAE}
\affiliation{Center for Astrophysics and Space Science (CASS), New York University Abu Dhabi, Abu Dhabi, UAE}

\author[0000-0003-0156-4564]{Luis Welbanks}
\affiliation{School of Earth and Space Exploration, Arizona State University, Tempe, AZ, USA}

\author[0000-0003-0973-8426]{Eva-Maria Ahrer}
\affiliation{Max Planck Institute for Astronomy, Königstuhl 17, D-69117 Heidelberg, Germany}

\author[0000-0002-4997-0847]{Duncan Christie}
\affiliation{Max Planck Institute for Astronomy, Königstuhl 17, D-69117 Heidelberg, Germany}

\author[0000-0002-2195-735X]{Louis-Philippe Coulombe}
\affiliation{Planétarium de Montréal, Espace pour la Vie, 4801 av. Pierre-de Coubertin, Montréal, Canada}
\affiliation{Institut Trottier de Recherche sur les Exoplanètes, 1375 Avenue Thérèse-Lavoie-Roux, Montréal, QC, H2V 0B3, Canada}

\author[0009-0007-1562-2944]{Gillis Lowry}
\affiliation{Carl Sagan Institute, Cornell University, 302 Space Sciences Building, Ithaca, NY 14853, USA}
\affiliation{Department of Physics \& Astronomy, San Francisco State University, 1600 Holloway Ave., San Francisco, CA 94132, USA}

\author[0000-0002-8517-8857]{Matthew M.\ Murphy}
\affiliation{Department of Physics and Astronomy, Michigan State University, East Lansing, MI, USA}

\author[0000-0002-9464-8101]{Adina D.\ Feinstein}
\affiliation{Department of Physics and Astronomy, Michigan State University, East Lansing, MI 48824, USA}

\author[0000-0002-6780-4252]{David Lafrenière}
\affiliation{Institut Trottier de Recherche sur les Exoplanètes, 1375 Avenue Thérèse-Lavoie-Roux, Montréal, QC, H2V 0B3, Canada}

\author[0000-0003-4816-3469]{Ryan J.\ MacDonald}
\affiliation{School of Physics \& Astronomy, University of St Andrews, North Haugh, St Andrews, KY16 9SS, UK}

\author[0000-0001-6707-4563]{Nathan J.\ Mayne}
\affiliation{Department of Physics and Astronomy, Faculty of Environment, Science and Economy, University of Exeter, Exeter EX4 4QL, UK}

\author[0000-0002-8163-4608]{Shang-Min Tsai}
\affiliation{Institute of Astronomy \& Astrophysics, Academia Sinica, Taipei 10617, Taiwan}

\author[0000-0002-9705-0535]{Maria Zamyatina}
\affiliation{Department of Physics and Astronomy, Faculty of Environment, Science and Economy, University of Exeter, Exeter EX4 4QL, UK}

\begin{abstract}
With its expanded wavelength coverage and increased precision compared to previous space-based observatories, JWST provides the opportunity to revisit benchmark planets and view them in a new light. Here, we conduct an in-depth study of the atmosphere of the hot-Saturn WASP-96\,b combining a new JWST NIRSpec/G395H transit with archival NIRISS/SOSS and VLT/FORS2 transmission spectra. The combined spectrum shows clearly-visible features from H$_2$O, CO$_2$, and Na. CO, though, remains unconstrained, precluding a firm metallicity derivation from free retrievals alone. However, self-consistent grids yield a broadly super-stellar atmospheric metallicity of 2--6$\times$ stellar. When combined with a roughly stellar C/O ratio ($0.41^{+0.10}_{-0.09}$ from self-consistent grids), we find that WASP-96\,b potentially formed via core-accretion beyond the H$_2$O snowline and subsequently accreted volatile-rich material. Free retrievals also find a moderate preference ($\ln B$=2.69) for models with SO$_2$ versus without. WASP-96\,b falls directly on the proposed ``SO$_2$ shoreline'' and the retrieved SO$_2$ abundance is well-matched to predictions from photochemical models. Our combined spectrum displays an optical slope, which our models fit with opacity from scattering aerosols --- either small-particle condensate clouds or photochemical hazes --- though we cannot completely rule out the broad wings of Na or the effects of stellar contamination. Future observations are necessary to disentangle these effects. Finally, we explore the possibility for limb asymmetry in WASP-96\,b's transmission spectrum and provide several tests to identify asymmetries in our data. We encourage the community to prioritize the development of a robust pathway to quantify the presence of limb asymmetry --- particularly for low signal-to-noise cases. 
\end{abstract}


\section{Introduction} 
\label{sec: Introduction}

The study of exoplanet atmospheres is now in an era of unprecedented observational sensitivity and precision, particularly due to JWST. Since the start of science operations in mid-2022, the community has leveraged the capabilities of JWST to access regimes not readily available to previous observatories: in-depth searches for atmospheres around rocky worlds \citep[e.g.,][]{greene_thermal_2023, moran_high_2023, zieba_no_2023, may_double_2023, lim_atmospheric_2023, zhang_gj_2024, radica_promise_2025} and investigations into the natures of sub-Neptunes \citep[e.g.,][]{madhusudhan_carbon-bearing_2023, kempton_reflective_2023, benneke_jwst_2024, piaulet-ghorayeb_jwstniriss_2024, ahrer_escaping_2025, davenport_toi-421_2025} being prime examples.

Although JWST has certainly opened up many new observational regimes ripe for exploration, it is also important to turn its powers to the atmospheres of giant planets. The high temperatures and large radii of hot-Jupiters and Saturns render their atmospheres easily observable \citep{seager_theoretical_2000, kempton_framework_2018}. Beyond sheer observability, though, transit spectroscopy of hot gas giants presents an incredible, high-S/N laboratory to study the myriad processes that govern planetary atmospheres: from photochemistry and other disequilibrium processes \citep{moses_disequilibrium_2011, tsai_photochemically_2023, welbanks_high_2024, fu_hydrogen_2024, mukherjee_effects_2025, crossfield_mapping_2025}, to atmosphere escape \citep{spake_helium_2018, mansfield_detection_2018, krishnamurthy_continuous_2025, allart_complex_2025}, cloud formation \citep{wakeford_high-temperature_2017, grant_jwst-tst_2023, dyrek_so2_2024}, and the impacts of planetary formation and migration \citep[e.g.,][]{oberg_effects_2011, madhusudhan_toward_2014, penzlin_bowie-align_2024, ahrer_tracing_2025}.

Recently, we have also begun to identify differences in the morning and evening limb transmission spectra of giant planets with JWST \citep[i.e., limb asymmetries;][]{espinoza_inhomogeneous_2024, murphy_evidence_2024, murphy_panchromatic_2025, mukherjee_cloudy_2025}. Although limb asymmetries have been detected at high resolution from the ground \citep[e.g.,][]{ehrenreich_nightside_2020, gandhi_spatially_2022}, such observations were out of reach for previous space-based instruments due to strong systematics and non-continuous observations. It is, therefore, still pertinent to revisit benchmark planets from the Hubble Space Telescope (HST) era to see what JWST can reveal about them.

WASP-96\,b, an inflated hot-Saturn \citep[M$\sim$0.48\,M$\rm _{Jup}$, R$\sim$1.2\,R$\rm _{Jup}$;][]{hellier_transiting_2014}, is one such benchmark world. It was the target of transit observations with VLT/FORS2 (0.36--0.82\,µm) by \citet{nikolov_absolute_2018}, as well as with the Wide Field Camera 3 (WFC3) instrument on HST (1.1--1.7\,µm) and the Spitzer Space Telescope's Infrared Array Camera (IRAC) by \citet{nikolov_solar--supersolar_2022}, leading to claims of an aerosol-free (i.e., cloud and haze free) terminator atmosphere --- mostly due to the highly broadened wings of the Na feature visible in the FORS2 spectrum \citep{nikolov_absolute_2018, nikolov_solar--supersolar_2022}. Although a followup Magellan/IMACS (0.44--0.97\,µm) transit spectrum by \citet{mcgruder_access_2022} did not show the broad Na wings as clearly as VLT/FORS2, and \citet{yip_compatibility_2021} highlighted the difficulties in combining space- and ground-based datasets, both studies also supported the aerosol-free interpretation of WASP-96\,b's transmission spectrum. \citet{nikolov_solar--supersolar_2022} also conducted an in-depth chemical analysis of WASP-96 itself, and combined with their transmission spectra found stellar-to-super-stellar abundances of O and Na in the atmosphere of WASP-96\,b.  

WASP-96\,b was then targeted with the Near Infrared Imager and Slitless Spectrograph \citep[NIRISS;][]{doyon_near_2023} on JWST as part of the Early Release Observations program. \citet{radica_awesome_2023} and \citet{taylor_awesome_2023} analyzed one transit using the Single Object Slitless Spectroscopy \citep[SOSS;][]{albert_near_2023} mode, covering wavelengths 0.6--2.85\,µm. They noted the presence of a short-wavelength slope in the continuous SOSS spectrum, which could potentially be masked by offsets when stitching together spectra from different instruments \citep{yip_compatibility_2021}. As demonstrated by previous work on model degeneracies \citep[e.g.,][]{Welbanks2019}, \citet{taylor_awesome_2023} found that the slope in the optical could be explained either by wings of a highly-broadened Na feature, or a scattering aerosol slope --- favouring the latter interpretation as general circulation models (GCMs) without aerosol opacity failed to provide an adequate fit to the data. \citet{taylor_awesome_2023} also concluded that the chemical composition of WASP-96\,b was roughly consistent with a solar metallicity atmosphere in chemical equilibrium, with potential indications of enhanced CO$_2$ and Na abundances (of roughly 10$\times$ solar). 

This apparent discrepancy between the solar H$_2$O abundances and super-solar CO$_2$ abundances was explored by \citet{rotman_enabling_2025}. Their analysis of the NIRISS spectrum of WASP-96\,b, using more flexible non-parametric models, suggest that the uncertainty on previously reported chemical abundances may be underestimated. The revised CO$_2$ abundance from \citet{rotman_enabling_2025} was consistent with expectations from solar and super-solar atmospheric metallicities.  

More recently, \citet{wang_comprehensive_2026} reanalyzed the NIRISS/SOSS transit and combined the resulting transmission spectrum with published VLT, HST, and Spitzer data. They revise the metallicity, via the H$_2$O abundance, downwards to sub-stellar values (assuming chemical equilibrium) and also find evidence for a grey cloud deck muting spectral features. In contrast to \citet{radica_awesome_2023} and \citet{taylor_awesome_2023} though, they do not find evidence for a possible scattering slope. \citet{wang_comprehensive_2026} also claim evidence for a $\sim$50\,s transit time offset in the Na feature, potentially indicating the presence of limb asymmetry.

In this work we conduct a thorough reinterpretation of WASP-96\,b's transmission spectra, including in the analysis a new transit with JWST's Near Infrared Spectrograph \citep[NIRSpec;][]{birkmann_near-infrared_2022}. This adds near-infrared (NIR) wavelength coverage from 3--5\,µm, providing access to critical molecules like CO$_2$, CO, and SO$_2$ and allowing for a more comprehensive dive into the atmospheric composition of this benchmark giant planet. We present our analysis as follows: Section~\ref{sec: Observations} outlines the observations and data analysis, and Section~\ref{sec: Modelling} the atmosphere models that we employ. We then describe our findings on the atmosphere composition of WASP-96\,b in Section~\ref{sec: Atmosphere Inferences}, as well as evidence for and implications of limb asymmetry in Section~\ref{sec: Limb Asymmetry}, before concluding in Section~\ref{sec: Conclusion}.

\section{Observations and Data Analysis} 
\label{sec: Observations}

\subsection{JWST NIRSpec/G395H}
\label{sec: NIRSpec}
We observed one transit of WASP-96\,b with JWST NIRSpec\footnote{GO 4082, PI: Radica} in BOTS mode \citep{birkmann_near-infrared_2022} using the G395H grating ($\sim$2.9 -- 5.0\,µm; R$\sim$3000). The time series observations (TSO) started on Aug 10, 2024 at 02:37 UTC and lasted 4.9\,hr (using 128 groups and 152 integrations), which captured the 2.4\,hr transit and 2.5\,hr of baseline observations before and after. 

We reduced the data with the widely-used \texttt{exoTEDRF} pipeline \citep{radica_exotedrf_2024, radica_awesome_2023, feinstein_early_2023}, closely following the procedures developed for NIRSpec/G395H observations in e.g., \citet{ahrer_escaping_2025, luque_insufficient_2025}. This includes subtracting a self-calibrated superbias frame created via an average of the first groups of every integration, group-level 1/$f$ correction, linearity correction, time-domain outlier rejection via the algorithm developed in \citet{radica_muted_2024}, and ramp fitting. We then re-subtract the background and any residual 1/$f$ noise after ramp fitting, interpolate bad pixels, and trace the spectrum on both the NRS1 and NRS2 detectors using the \texttt{edgetrigger} algorithm \citep{radica_applesoss_2022}, which is well suited to the curved traces of various instruments \citep[e.g.,][]{radica_constraining_2025}. 

We also apply the principal component analysis (PCA) step commonly employed for SOSS observations \citep{coulombe_broadband_2023, radica_muted_2024, coulombe_highly_2025}, and recently for MIRI \citep{luque_insufficient_2025, connors_uniform_2025}, to identify potential detector-correlated trends in the data. This analysis identifies a sub-pixel y-position drift on both detectors over the course of the time series\footnote{Several diagnostic plots from the data analysis including the PCA outputs are included in this Zenodo repository: \url{https://zenodo.org/records/17065171}. Hereafter, whenever a Zenodo repository is referred to, it is always this one.}. However, instead of just using this eigenvalue timeseries as a light curve detrending vector, we use the power of PCA to reconstruct the entire TSO dataset removing this component. This novel analysis step is reminiscent of techniques used to remove telluric contamination in high-resolution observations \citep[e.g.,][]{pelletier_where_2021, pelletier_vanadium_2023}, and ensures that the undesired detector-correlated noise is entirely removed from the data prior to light curve fitting. Of course, this analysis can only remove noise sources correlated with detector trends, meaning that any other source of \textit{astrophysical} noise will still be present in the light curves. We find that removing this sub-pixel y-position drift from the observations results in a modest reduction of the white light curve scatter ($\sim$10\,ppm, or $\sim$7\%). 

Finally, we extract the stellar spectra using a simple box aperture with a full-width of eight pixels. We explore tweaking several of these reduction steps (e.g., using the default STScI superbias reference file, optimal extraction) and find negligible difference in the end transmission spectrum.

\subsection{JWST NIRISS/SOSS}
\label{sec: NIRISS}
We combine the NIRSpec observations with archival NIRISS/SOSS observations taken as part of the JWST Early Release Observations program. These data were originally presented in \citet{radica_awesome_2023} with further analysis in \citet{taylor_awesome_2023}. However, we believe it prudent to re-reduce the raw data, thereby allowing us to incorporate the approximately two years of advances in JWST data analysis since the original publication. 

To that end, we again apply the \texttt{exoTEDRF} pipeline to these data, closely following the steps laid out in \citet{radica_muted_2024} and \citet{radica_promise_2025}. Notable changes to the analysis compared to that of \citet{radica_awesome_2023} include separate pre- and post-step background scaling \citep{lim_atmospheric_2023, fournier-tondreau_near-infrared_2024}, use of the time-domain outlier rejection instead of the standard up-the-ramp flagging offered by the STScI pipeline, more robust masking of background contaminants during the group-level 1/$f$ correction, and the application of the PCA reconstruction step described above. For this, we remove three components identified by the PCA: a sub-pixel y-position drift, the beating pattern linked to the telescope thermal control \citep{albert_near_2023}, as well as the minor tilt event noted by \citet{radica_awesome_2023}. We also extract the stellar spectra using a simple box aperture with a width of 30 pixels, which we find minimizes the white light curve scatter, instead of the \texttt{ATOCA} algorithm \citep{darveau-bernier_atoca_2022} since \citet{radica_awesome_2023} demonstrated that the effects of the SOSS order self-contamination is indeed negligible for this dataset. 

Finally, we apply the post processing steps described in Section 2.2.4 of \citet{radica_awesome_2023}, the efficacy of which was recently validated by \citet{rotman_enabling_2025}, to undo the dilution caused by two background contaminants that intersect the target trace. 

\subsection{Refining WASP-96b's Orbital Solution}
\label{sec: Joint Fits}

\begin{deluxetable}{c|cc}
 \centering
 \tabletypesize{\footnotesize}
 \label{tab: WLC Parameters}
 \tablecaption{Best-fitting transit parameters from the joint transit and RV fit}
 \tablehead{Parameter & Prior Range & Value}
    \startdata
     Per [d] & $\mathcal{N}$[3.42525674, 0.1] & 3.4252564$^{+0.0000002}_{-0.0000002}$ \\
     T0$\rm _{SOSS}$ [MJD] & $\mathcal{U}$[T0$\pm$2\,hr] & 59751.32470$^{+0.00003}_{-0.00003}$ \\
     T0$\rm _{NRS}$ [MJD] & $\mathcal{U}$[T0$\pm$2\,hr] & 60532.28316$^{+0.00003}_{-0.00003}$ \\
     Rp/Rs$\,\rm _{TESS}$ & $\mathcal{U}$[0.01, 0.9] & 0.1172$^{+0.0006}_{-0.0009}$ \\
     Rp/Rs$\,\rm _{SOSS1}$ & $\mathcal{U}$[0.01, 0.9] & 0.1197$^{+0.0004}_{-0.0004}$ \\
     Rp/Rs$\,\rm _{SOSS2}$ & $\mathcal{U}$[0.01, 0.9] & 0.1204$^{+0.0005}_{-0.0005}$ \\
     Rp/Rs$\,\rm _{NRS1}$ & $\mathcal{U}$[0.01, 0.9] & 0.1188$^{+0.0002}_{-0.0002}$ \\
     Rp/Rs$\,\rm _{NRS2}$ & $\mathcal{U}$[0.01, 0.9] & 0.1192$^{+0.0004}_{-0.0005}$ \\
     a/Rs & $\mathcal{U}$[1, 25] & 8.9890$^{+0.0222}_{-0.0216}$ \\
     b & $\mathcal{U}$[0, 1] & 0.7298$^{+0.0020}_{-0.0020}$ \\
     q1$\rm _{TESS}$ & $\mathcal{U}$[0, 1] & 0.34$^{+0.05}_{-0.05}$ \\
     q2$\rm _{TESS}$ & $\mathcal{U}$[0, 1] & 0.14$^{+0.10}_{-0.18}$ \\
     q1$\rm _{SOSS1}$ & $\mathcal{U}$[0, 1] & 0.14$^{+0.01}_{-0.01}$ \\
     q2$\rm _{SOSS1}$ & $\mathcal{U}$[0, 1] & 0.46$^{+0.13}_{-0.14}$ \\
     q1$\rm _{SOSS2}$ & $\mathcal{U}$[0, 1] & 0.29$^{+0.02}_{-0.02}$ \\
     q2$\rm _{SOSS2}$ & $\mathcal{U}$[0, 1] & 0.58$^{+0.10}_{-0.10}$ \\
     q1$\rm _{NRS1}$ & $\mathcal{U}$[0, 1] & 0.05$^{+0.01}_{-0.01}$ \\
     q2$\rm _{NRS1}$ & $\mathcal{U}$[0, 1] & 0.11$^{+0.08}_{-0.15}$ \\
     q1$\rm _{NRS2}$ & $\mathcal{U}$[0, 1] & 0.04$^{+0.01}_{-0.01}$ \\
     q2$\rm _{NRS2}$ & $\mathcal{U}$[0, 1] & 0.35$^{+0.23}_{-0.29}$ \\
    \enddata
    \tablecomments{$\rm q1$ and $\rm q2$ refer to the \citet{kipping_efficient_2013} re-parameterization of the quadratic limb-darkening law.
    $T_0$ is fit as the NIRISS mid-transit time and also propagated here to the epoch of the NIRSpec transit.}
\end{deluxetable}

\begin{figure*}
    \centering
    \includegraphics[width=0.9\linewidth]{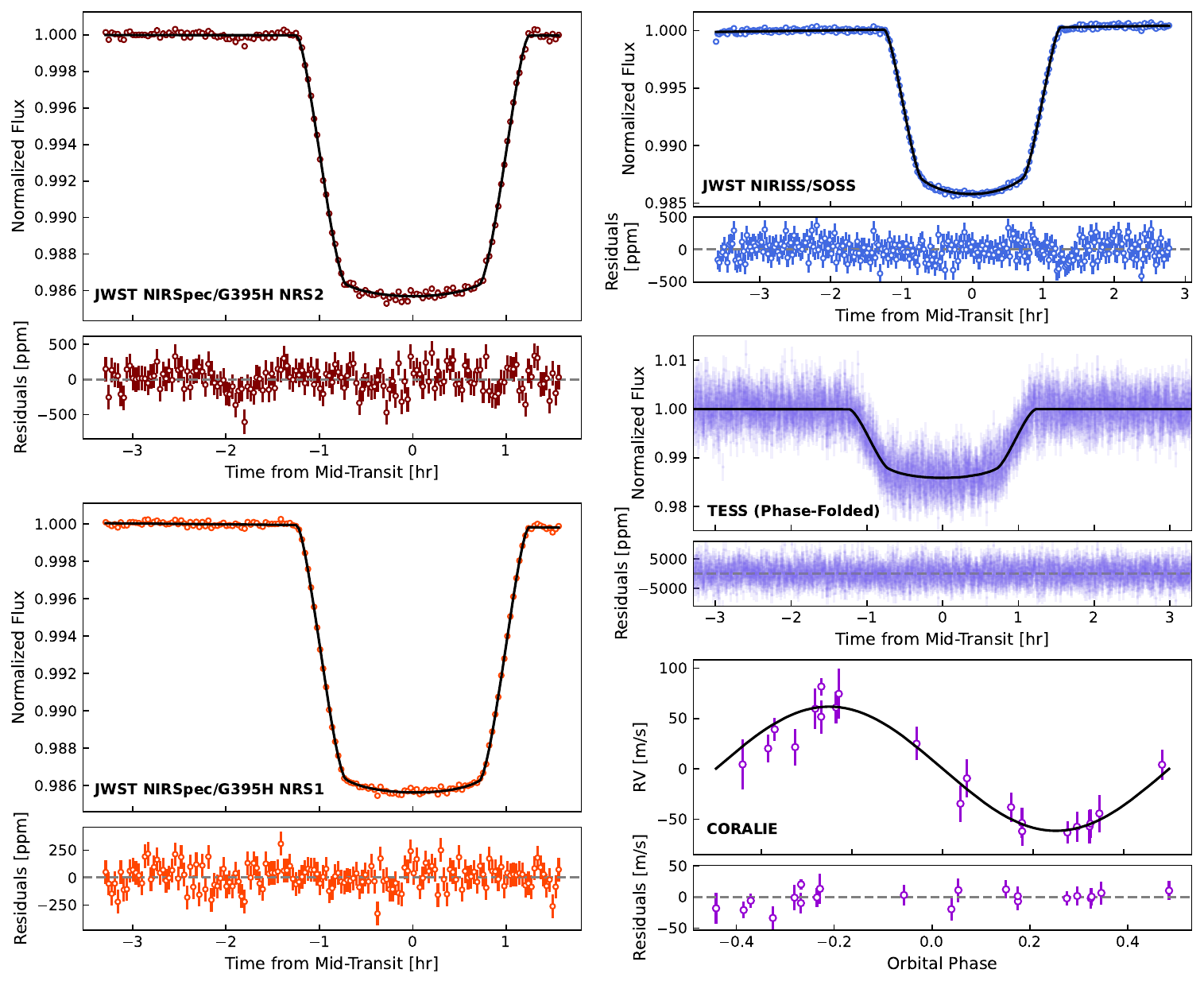}
    \caption{Data (coloured points) and best fitting models (black) from the joint transit and RV fit. Residuals to the best-fitting model are shown below each dataset. The TESS and CORALIE data have been phase folded to the best-fitting orbital period.}
    \label{fig:JointFitResults}
\end{figure*}

A goal of this study is to identify and characterize potential limb differences in WASP-96\,b's atmosphere. Since limb differences can be highly degenerate with the transit parameters themselves \citep[e.g., T$_0$;][]{espinoza_constraining_2021, murphy_analytic_2024, fu_overcast_2025}, it is imperative to obtain the most accurate possible orbital solution for WASP-96\,b before embarking on this endeavour. 

To this end, we supplement our four JWST light curves (two NIRISS/SOSS orders and two NIRSpec detectors) with publicly available light curves from the Transiting Exoplanet Survey Satellite \citep[TESS;][]{ricker_transiting_2014}, and CORALIE radial velocity (RV) measurements from \citet{hellier_transiting_2014}. For TESS, we use the Pre-search Data Conditioning Simple Aperture Photometry \citep[PDCSAP;][]{Jenkins_tess_2016} products from sectors 2 and 29 available from the MAST archive. We do not include the HST light curves from \citet{nikolov_solar--supersolar_2022} as their large gaps render them non-ideal for constraining orbital properties, nor the VLT or Magellan light curves as they are not publicly available. We convert all timestamps to MJD using routines in the \texttt{astropy.Time} library.

We use the flexible \texttt{juliet} package \citep{espinoza_juliet_2019} to jointly fit the transit and RV datasets. The transit model uses \texttt{batman} \citep{kreidberg_batman_2015}, assuming a circular orbit \citep{hellier_transiting_2014} and with the orbital period, $P$, mid transit time (fixed to be that of the NIRISS/SOSS transit), $T_0$, scaled planet radius, $R_p/R_*$, impact parameter, $b$, scaled semi-major axis, $a/R_*$, and the two parameters of the \citet{kipping_efficient_2013} parameterization of the quadratic limb-darkening law, $q_1$ and $q_2$, as free parameters. For each JWST dataset, we also include a linear slope with time as a systematics model. The RV fit calls the Keplerian solution from \texttt{radvel} \citep{fulton_radvel_2018}, with the RV offset and semi-amplitude as additional free parameters. The orbital parameters (i.e., $P$, $T_0$, $a/R_*$, $b$) are shared between all data, and the rest fit individually to a given dataset as appropriate. An error inflation term is also fit to each dataset, added in quadrature to the errors such that the final reduced $\chi^2$ of the fit is unity. All parameters use wide, uninformative priors except the period to which we give a Gaussian prior based off of the transit timing analysis of \citet{kokori_exoclock_2023}. Priors and best-fitting values for relevant parameters are included in Table~\ref{tab: WLC Parameters} and the best-fitting models for each dataset are shown in Figure~\ref{fig:JointFitResults}.

Our refined orbital parameters are generally consistent with those previously presented in the literature \citep[e.g.,][]{hellier_transiting_2014, nikolov_solar--supersolar_2022, patel_empirical_2022, mcgruder_access_2022}. In particular our scaled semi-major axis and impact parameter are consistent within errors to the values derived by \citet{nikolov_solar--supersolar_2022} using HST, Spitzer, TESS, and ground-based data, as well as \citet{patel_empirical_2022} using TESS. However, our precision is higher due in large part to the JWST transits enabling precise constraints on the transit duration from which these parameters are derived \citep[e.g.,][]{seager_unique_2003, carter_benchmark_2024}. The consistency in the period is unsurprising given that it is primarily constrained by TESS data which was used in all analyses mentioned above.

\subsection{JWST Light Curve Fits}
\label{sec: Light Curve Fits}

\begin{figure*}
    \centering
    \includegraphics[width=0.95\linewidth]{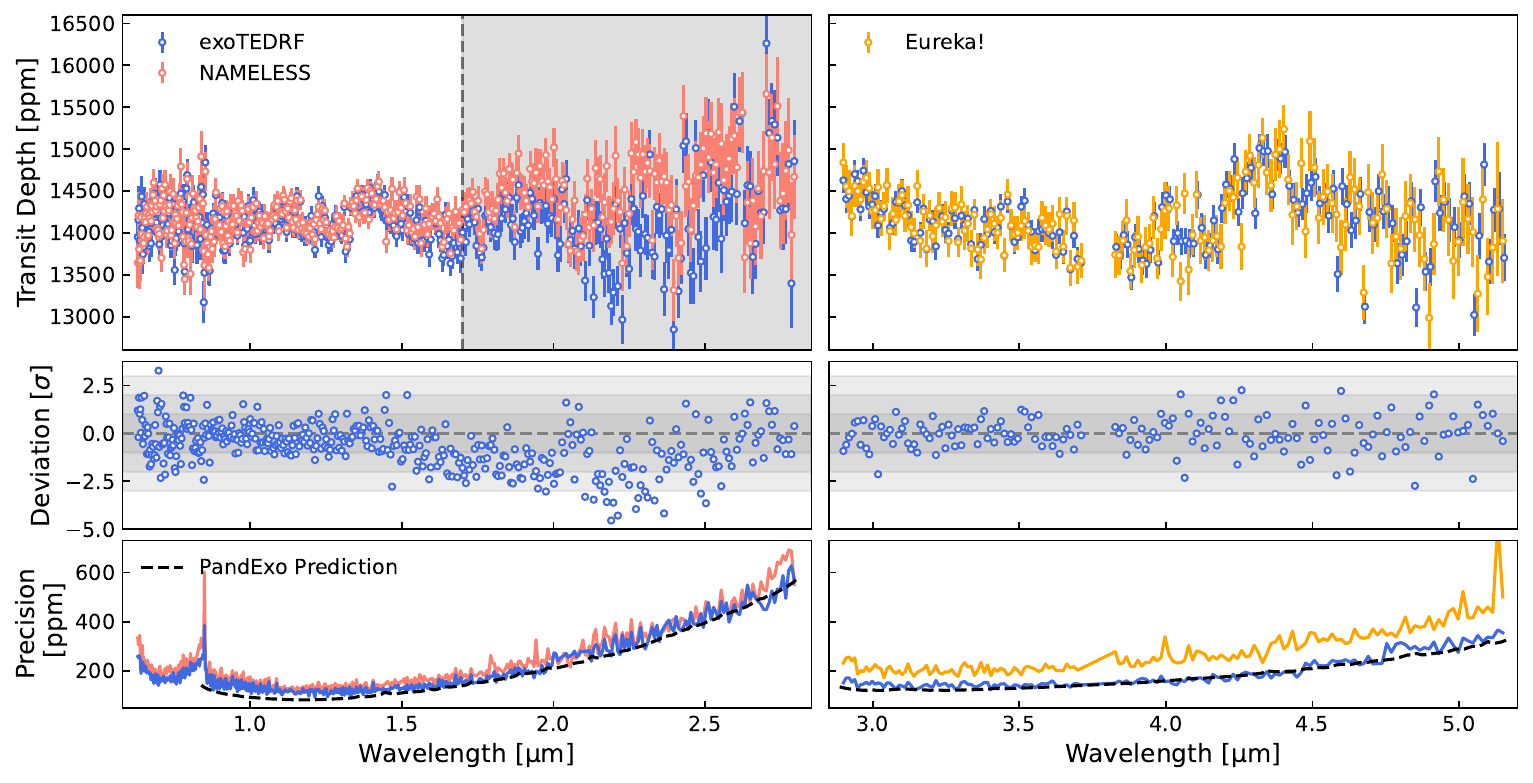}
    \caption{Comparison between our nominal \texttt{exoTEDRF} spectra and alternate reductions with \texttt{NAMELESS} for NIRISS (left panels) and \texttt{Eureka!}\ for NIRSpec (right panels).
    \emph{Top}: The two spectra produced for each instrument overplotted. The grey shading in the NIRISS panel denotes wavelengths not used in the comparative retrievals (see Section~\ref{sec: Modelling}).
    \emph{Middle}: Error-normalized differences for each instrument. There is a significant divergence between the two NIRISS/SOSS spectra redwards of $\sim$1.7\,µm which can be attributed to differences in 1/$f$ noise correction methodologies (see Appendix~\ref{app: SOSS 1/f}).
    \emph{Bottom}: Light curve scatter as a function of wavelength compared to \texttt{PandExo} predictions.}
    \label{fig:Spectra_Compare_R300}
\end{figure*}

We then fit the spectrophotometric light curves from the JWST datasets, fixing the orbital parameters to those from Table~\ref{tab: WLC Parameters}. We use the \texttt{exoUPRF} package \citep{radica_exouprf_2024, ahrer_escaping_2025} for these, and fit the light curves for each instrument at three different resolutions: $R$=100, $R$=300, and the native detector resolution ($R$$\sim$3500 for NIRSpec and $\sim$700 for NIRISS). For each light curve, we fit the scaled planet radius, the transit zero point, a linear slope, and the additive error inflation term. We experiment with a number of limb darkening treatments, including varying the parameterization (quadratic, \citet{kipping_efficient_2013} quadratic, four-parameter), as well as the degree of flexibility (freely fitted, fixed, prior) and find negligible differences in the transmission spectra at all wavelengths. We therefore elect to use the quadratic law and put Gaussian priors on the limb darkening coefficients centered on the values predicted by \texttt{ExoTiC-LD} \citep{Grant2024ExoTiC-LD:Coefficients} using the 3D \texttt{stagger} stellar grid \citep{magic_stagger-grid_2015}, and with widths of 0.2 as recommended by \citet{patel_empirical_2022}.

The final transmission spectra at $R$=300 are shown in Figure~\ref{fig:Spectra_Compare_R300} for both instruments. Other comparisons and light curve fitting diagnostics are included in the Zenodo repository. We also perform an additional analysis of both datasets using \texttt{NAMELESS} \citep{coulombe_broadband_2023,coulombe_highly_2025} for NIRISS and \texttt{Eureka!} \citep{bell_eureka_2022} for NIRSpec. Details of these reductions can be found in Appendix~\ref{app: additional reductions}, and the final $R$=300 spectra are compared to the nominal \texttt{exoTEDRF} spectra in Figure~\ref{fig:Spectra_Compare_R300}.

\subsubsection{Asymmetric Fits}
\label{sec: Asymetric Fits}

\begin{figure*}
    \centering
    \includegraphics[width=0.95\linewidth]{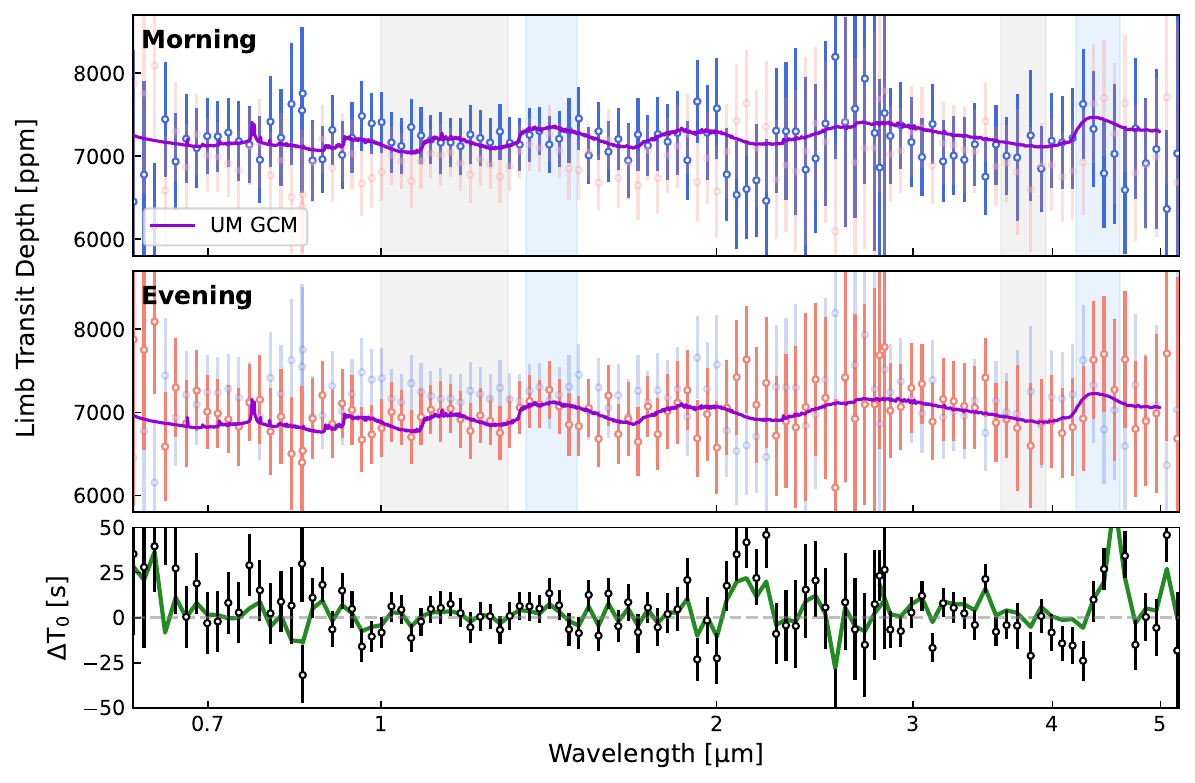}
    \caption{WASP-96\,b's morning and evening limb transmission spectra as observed with JWST.
    \emph{Top}: The morning-limb transmission spectrum (blue data points) compared to the evening-limb spectrum (faded red). Overplotted in purple is the morning-limb spectrum from the aerosol-free, 10$\times$ solar UM GCM run (see Appendix~\ref{app: GCM}). Blue and grey shaded rectangles denote the in-band and out-of-band wavelengths, respectively, for the H$_2$O and CO$_2$ band amplitude calculations (see Section~\ref{sec: Limb Holistic}).
    \emph{Middle}: Inverse of the above, focusing on the evening-limb spectrum. 
    \emph{Bottom}: Fitting mid-transit time as a function of wavelength assuming a uniform-limb (i.e., \texttt{batman}) planet (black points). In green is the $\rm T_0$ spectrum derived from the asymmetric \texttt{catwoman} fits using the formalism of \citet{murphy_analytic_2024}.}
    \label{fig:Limb_Spectra}
\end{figure*}

Finally, we attempt to extract signatures of inhomogeneous morning and evening limbs on WASP-96\,b via the transit light curve. We use the \texttt{catwoman} \citep{jones_catwoman_2020, espinoza_constraining_2021} package for this, which assumes that a transiting planet can be approximated as two conjoined semi-circles which can have different radii --- thereby allowing for the determination of independent morning and evening limb spectra \citep{von_paris_inferring_2016, kempton_framework_2018, espinoza_constraining_2021}.

Again, we fix the orbital parameters to those from Table~\ref{tab: WLC Parameters}. We then fix the limb-darkening to the quadratic law predictions from \texttt{ExoTiC-LD} and include a linear slope as a systematics model. T$_0$, in particular, is degenerate with the presence of limb asymmetry, however, our sub-10\,s precision on $T_0$ should allow for the extraction of limb asymmetries of a few scale heights \citep{murphy_analytic_2024}. We fit the light curves at three resolutions: $R$=100, 50, and 25, and the morning and evening limb spectra at $R$=50 are shown in Figure~\ref{fig:Limb_Spectra}. In Section~\ref{sec: Limb Asymmetry} we conduct a series of tests to quantify the degree to which the morning and evening spectra differ and thus identify the presence, or lack thereof, of limb asymmetry in our observations.

\section{Atmosphere Modelling} 
\label{sec: Modelling}

To interpret the atmospheric spectra of WASP-96\,b we use several Bayesian retrieval and forward modelling codes (\texttt{POSEIDON}, \texttt{NemesisPy}, \texttt{PyratBay}, \texttt{Aurora}, \texttt{ScCHIMERA}). This allows us to ensure that our results are robust to the particularities of a given retrieval code. We also explore the impacts of different retrieval setups (e.g., free vs.\ chemically consistent vs.\ radiative-convective-equilibrium models). We include the optical VLT/FORS2 transmission spectrum from \citet{nikolov_absolute_2018} in our modelling to extend the wavelength coverage to $\sim$0.35\,µm. We also test adding the Magellan/IMACS, and HST/WFC3 spectra, from \citet{mcgruder_access_2022} and \citet{nikolov_solar--supersolar_2022} respectively, but find their impacts to be negligible. This is due to these spectra providing redundant wavelength coverage at lower precision than VLT/FORS2 and NIRISS/SOSS, respectively.   

The base setup for each modelling code is described in the following sections. An example corner plot from the \texttt{POSEIDON} free retrieval on the nominal  VLT + $R$=300 JWST NIRISS \& NIRSpec data combination is included in Appendix~\ref{app: example corner} and corner plots from other relevant runs are included in the associated Zenodo repository.  

\begin{figure*}
    \centering
    \includegraphics[width=0.95\linewidth]{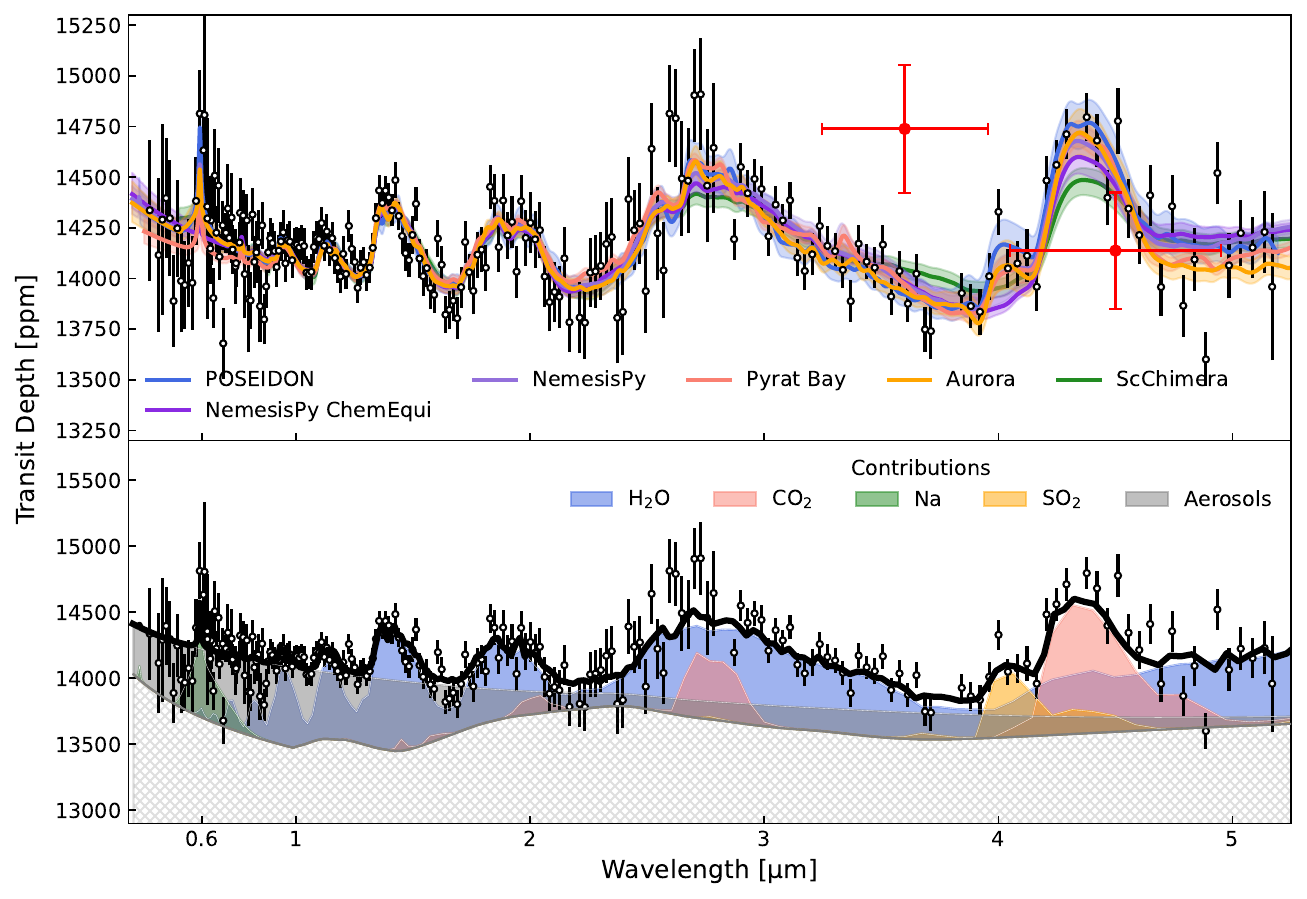}
    \caption{Results of modelling WASP-96\,b's transmission spectrum.
    \emph{Top}: Best-fitting atmosphere models from each retrieval code (coloured lines) along with the 2-$\sigma$ confidence envelopes (coloured shading) overplotted on the combined ground-based + JWST spectrum (black data points). The JWST data have been binned from the nominal resolution of $R=300$ to $R\sim100$ for plotting purposes. Also shown in red, but not included in the retrievals, are the Spitzer 3.6 and 4.5\,µm transit depths from \citet{nikolov_solar--supersolar_2022}. The 3.6\,µm point in particular, is discrepant with the NIRSpec observations. 
    \emph{Bottom}: Spectral decomposition of the transmission spectrum to show contributions from various chemical species as well as aerosols (here primarily a scattering slope).
    }
    \label{fig:Spectrum_Decomp}
\end{figure*}

\subsection{\texttt{POSEIDON}}
\label{sec: poseidon}

The first code that we use is the open-source \texttt{POSEIDON} package \citep{macdonald_hd_2017, macdonald_poseidon_2023}. We run ``free'' retrievals, where each absorber can vary independently and has a constant abundance with altitude. We include opacity from H$_2$O, CO$_2$, CO, CH$_4$, H$_2$S, SO$_2$, Na, K, HCN, and NH$_3$ (all references included in Table~\ref{tab: Opacities}), as well as H$_2$-H$_2$ and H$_2$-He collision-induced absorption (CIA). We include contributions of aerosols via a ``cloud-haze" prescription, using a grey slab cloud at pressure $P_{\rm cloud}$ and a modified Rayleigh scattering slope, with an enhancement factor $\alpha$ and scattering slope $\gamma$, where $\gamma=-4$ represents H$_2$ Rayleigh scattering \citep[e.g.,][]{macdonald_hd_2017, pinhas_retrieval_2018}. We also allow for inhomogeneous ``patchy'' aerosols via the prescription of \citet{LineParmentier2016-patchy}.
 
We generate an isothermal, plane-parallel atmosphere model spanning 2 to $-$7\,bar in log pressure at a resolution of $R$=20000, which has been demonstrated to be sufficient for JWST transmission spectra of giant planets \citep[e.g.,][]{louie_jwst-tst_2025}. We include offsets between each dataset (relative to NIRISS) when jointly retrieved on, and sample the posterior space using the \texttt{MultiNest} sampling algorithm \citep{feroz_multinest_2009} with 1000 live points.

\subsection{\texttt{PyratBay}}
\label{sec: Pyrat Bay}

We also used the \texttt{PyratBay} framework \citep{CubillosBlecic2021} to model the atmosphere of WASP-96\,b. \texttt{PyratBay} is a comprehensive spectrum synthesis and atmospheric retrieval tool designed for exoplanet studies. For this analysis, we employed a free-chemistry retrieval framework and generated models incorporating the NIRISS and NIRSpec observations, with and without ground-based data from VLT/FORS2. We further examined how different data resolutions affect the retrieved atmospheric properties and overall conclusions. 

We include opacity from H$_2$O, Na, K, CH$_4$, NH$_3$, HCN, CO, CO$_2$, and SO$_2$ (summarized in Table~\ref{tab: Opacities}). In addition, we incorporated the Rayleigh scattering from \citet{LecavelierEtal2008aaRayleighHD189733b} which includes the strength ($f_{\rm ray}$) and power-law index ($\alpha_{\rm ray}$) of the scattering opacity cross section, and a grey cloud deck where the atmosphere becomes instantly opaque at all wavelengths at a requested pressure level. Inhomogeneous clouds were incorporated following \citet{LineParmentier2016-patchy}. We also adopt the parametric temperature–pressure profile of \citet{madhusudhan_temperature_2009}.

The set of free parameters included the abundances of the nine aforementioned chemical species assuming uniform priors, six parameters describing the temperature–pressure profile, one parameter for the planetary radius at a reference pressure of 0.1 bar, two parameters for Rayleigh scattering, one parameter for the cloud-top pressure of a grey cloud model, and one parameter describing cloud patchiness ($f_\mathrm{patchy}$). The parameter space was explored using the Nested Sampling algorithm implemented in \texttt{PyMultiNest} \citep{skilling_nested_2006,buchner_x-ray_2014}, with 2,000 live points in all retrievals.

\subsection{\texttt{NemesisPy}}

\texttt{NEMESIS} is a radiative transfer and retrieval tool originally developed to study planetary atmospheres within the solar system \citep{Irwin2008}, but extensively adapted to study exoplanet atmospheres \citep[e.g.][]{Barstow2017,Taylor2023}. The framework is now fully pythonized, and we use this version of the code for this analysis \citep[called \texttt{NemesisPy} from here on out;][]{Yang2024}. \texttt{NemesisPy} uses the correlated-k method to compute the molecular and atomic opacities \citep{Lacis1991} and nested sampling, specifically \texttt{PyMultiNest} to sample the parameter-space \citep{skilling_nested_2006,buchner_x-ray_2014}. Included opacity sources are the same as above and summarized in Table~\ref{tab: Opacities}. Gas opacities are computed using k-tables with resolution R=1000, obtained from the ExoMol database \citep{Chubb2021}, before being channel averaged to the resolution of the observations. We perform both a free chemistry and chemical equilibrium retrieval in which we utilize \texttt{FastChem} \citep{stock_fastchem_2022} to obtain values for the atmospheric metallicity and C/O. 

We model clouds/hazes following the prescription derived in \citet{macdonald_hd_2017}, where we fit for a cloud top pressure, a Rayleigh enhancement factor, a scattering slope, and a cloud fraction. As is typical when combining observations from different instruments, we anchor our retrieval to the NIRISS/SOSS observations and fit for offsets for VLT, Magellan, NIRSpec NRS1 and NRS2. The priors used for all parameters are listed in Table \ref{tab: Atmosphere Inferences}. We sample the parameter space using 1000 live points and have an evidence tolerance of 0.5.

\subsection{\texttt{Aurora}}

The final retrieval framework that we use is \texttt{Aurora} \citep{welbanks_aurora_2021}. \texttt{Aurora} assumes a one-dimensional plane-parallel atmosphere and solves the requisite radiative transfer equations for hydrostatic equilibrium to produce a transmission spectrum of the planet.

As with the other retrieval frameworks, we model the temperature-pressure profile using the six-parameter prescription of \citet{madhusudhan_temperature_2009}. We use the two sector cloud and haze prescription from \citet{welbanks_aurora_2021} where the hazes are parametrized as a deviation from H$_2$-Rayleigh scattering \citep{LecavelierEtal2008aaRayleighHD189733b} and clouds are parametrized through a grey cloud-deck at a given pressure layer. The clear and cloudy/hazy atmospheric models are combined following \citet{line_influence_2016}. 

Our atmospheric models consider opacity from the same set of species as the other codes (summarized in Table~\ref{tab: Opacities}), and fit for offsets between instruments and detectors, as has been previously done for panchromatic JWST spectra \citep[e.g.,][]{carter_benchmark_2024}. We apply a nested sampling approach for our retrievals, using the MultiNest sampler \citep{feroz_multinest_2009} via the PyMultiNest python wrapper \citep{buchner_x-ray_2014}, with 1000 live points used for sampling.

\subsection{\texttt{ScCHIMERA}}

We further consider atmospheric models under the assumption of radiative-convective-thermo/photo-chemical equilibrium. These models are generated using \texttt{ScCHIMERA}  \citep{bell_methane_2023, welbanks_high_2024,wiser_precise_2025} calculating the vertical temperature structure and chemical composition of the planet's atmosphere for a given heat redistribution, atmospheric metallicity, and C/O ratio. ScCHIMERA solves the radiative transfer between layers utilizing the two-stream approximation \citep{Toon1989} and absorbers expected in exoplanet atmospheres under the assumption of thermochemical equilibrium; these are calculated using the \texttt{CEA2} module \citep{GordonMcbride1994}, which, for each layer, minimizes the Gibbs free energy and provides chemical abundances. A description of the thermochemical data sources is provided in \citet{mcbride_thermodynamic_1993}, with the iteration done using a Newton-Raphson scheme \citep{McKay1989}. Then, we solve for chemical kinetics in the atmosphere given an incident stellar flux to account for the effects from chemical disequilibrium (e.g., photochemistry) using \texttt{Photochem} \citep{wogan_open-source_2025}. Our grid of models follows the same spacing and dimensions as that used in \citet{radica_awesome_2023} with the exception of metallicity, which in this work has log-metallicities between $-1.0$ and 1.625 in spacing of 0.125~dex. We adopt solar abundances from \citet{2009LanB...4B..712L}. 

We use this grid of models to fit the transmission of WASP-96\,b using \texttt{CHIMERA} \citep{Line_2013} considering absorption due to H$_2$O, CO,  CO$_2$, CH$_4$, NH$_3$, H$_2$S, HCN, C$_2$H$_2$, Na,  K, and SO$_2$ as well as H$_2$-H$_2$ and H$_2$-He CIA (references in Table~\ref{tab: Opacities}). We include the presence of inhomogeneous clouds/hazes following the same description as in Aurora above. In total, we fit for 12 parameters: redistribution factor $f$, atmospheric metallicity $Z$, C/O, radius scaling $\times R_p$, cloud opacity $\kappa$, Rayleigh scattering amplitude $a$, haze slope $\gamma$, cloud/haze covering fraction $\phi_{\rm cloud\,and \, hazes}$, and offsets for all instruments relative to NIRISS. Parameter estimation is performed using MultiNest \citep{feroz_multinest_2009} via the PyMultinest wrapper \citep{buchner_x-ray_2014}.

\section{Optical-to-IR Atmosphere Characterization of WASP-96 {\MakeLowercase b}}
\label{sec: Atmosphere Inferences}

\subsection{Overview of Model Results}
\label{sec: Overview of Results}

Table~\ref{tab: Atmosphere Inferences} presents the results from the full gamut of retrieval tests run on the transmission spectra of WASP-96\,b. We take as our nominal results the inferences from the full $R$=300 JWST + VLT combination. We also run a resolution test using \texttt{Pyrat Bay} varying the resolution of the JWST data, and another test with \texttt{NemesisPy} including the Magellan and HST data in the analysis, neither of which significantly affects the results. The best fitting models from each retrieval code on the nominal data configuration are shown in Figure~\ref{fig:Spectrum_Decomp}. We also include, in the figure and not the retrievals, the 3.6 and 4.5\,µm \textit{Spitzer}/IRAC transit depths from \citet{nikolov_solar--supersolar_2022} for comparison. Although the 4.5\,µm point agrees well with the new JWST NIRSpec data, the 3.6\,µm point does not --- making this one of the few times that JWST data have contradicted previous observations from \textit{Spitzer}.

There are significant differences between the NIRISS spectra from \texttt{exoTEDRF} and \texttt{NAMELESS} at wavelengths $\gtrsim$2\,µm (Figure~\ref{fig:Spectra_Compare_R300}), which we show in Appendix~\ref{app: SOSS 1/f} is due to the 1/$f$-noise correction method and argue that the group-level correction used in \texttt{exoTEDRF} is more correct. However, we verify that our atmosphere inferences still hold when retrieving on the \texttt{NAMELESS} + \texttt{Eureka!} spectrum when removing this discrepant region.

In the following sections, we highlight some key findings and their implications for the atmosphere of WASP-96\,b. Where comparisons with stellar values are made, stellar abundances are sourced from \citet{nikolov_solar--supersolar_2022}. We also use \citet{asplund2009} as the reference for solar values to remain consistent with \citet{nikolov_solar--supersolar_2022}.

\movetabledown=6cm
\begin{rotatetable*}
\begin{deluxetable*}{cc||cc|ccc|cccc|c|c}
 \tabletypesize{\scriptsize}
 \tablecaption{Retrieved atmosphere constraints for WASP-96\,b}  
 \label{tab: Atmosphere Inferences}
 \tablehead{
  \multirow{2}{*}{Parameter} & \multirow{2}{*}{Prior Range} & \multicolumn{2}{c}{\texttt{POSEIDON}} & \multicolumn{3}{c}{\texttt{NemesisPy}} & \multicolumn{4}{c}{\texttt{Pyrat Bay}} & \texttt{Aurora} & \texttt{ScCHIMERA}\\
  & & Free$^{a}$ & Alt.\ Free$^{b}$ & Free$^{a}$ & Alt.\ Free$^{c}$ & Chem.~Equi.$^{a}$ & Free$^{a}$ & $R=100^{d}$ & $R=300^{d}$ & pixel$^{d}$ & Free & Grid
           }
    \startdata
     T [K]$^{e}$ & $\mathcal{U}$[200, 2000] & $1093^{+54}_{-49}$ & $1044^{+131}_{-79}$ & $1017^{+79}_{-72}$ & $1085^{+66}_{-64}$ & $961^{+46}_{-37}$ & $1029^{+94}_{-82}$ & $977^{+96}_{-84}$ & $1036^{+215}_{-119}$ & $949^{+102}_{-106}$ & $939^{+25}_{-27}$ & - \\
     $\rm \log H_2O$ & $\mathcal{U}$[$-12$, $-1$] & $-2.61^{+0.27}_{-0.30}$ & $-2.82^{+0.62}_{-0.76}$ & $-2.82^{+0.24}_{-0.26}$ & $-2.79^{+0.25}_{-0.26}$ & - & $-2.67^{+0.29}_{-0.28}$ & $-2.30^{+0.38}_{-0.41}$ & $-2.36^{+0.32}_{-0.38}$ & $-2.63^{+0.28}_{-0.26}$ & $-2.76^{+0.30}_{-0.28}$ & - \\    
     $\rm \log CH_4$ & $\mathcal{U}$[$-12$, $-1$]  & $<-5.91$ & $<-5.35$ & $<-5.81$ & $<-5.88$ & - & $-5.75^{+0.27}_{-0.26}$ & $-6.58^{+0.72}_{-2.73}$ & $-5.87^{+0.35}_{-0.42}$ & $-5.93^{+0.26}_{-0.26}$ & $<-6.18$ & - \\
     $\rm \log CO_2$ & $\mathcal{U}$[$-12$, $-1$]  & $-4.65^{+0.34}_{-0.39}$ & $-4.15^{+0.70}_{-0.77}$ & $-4.91^{+0.30}_{-0.30}$ & $-4.88^{+0.31}_{-0.31}$ & - & $-4.48^{+0.32}_{-0.30}$ & $-4.49^{+0.42}_{-0.43}$ & $-4.34^{+0.35}_{-0.38}$ & $-4.50^{+0.31}_{-0.30}$ & $-4.46^{+0.34}_{-0.31}$ & - \\
     $\rm \log CO$ & $\mathcal{U}$[$-12$, $-1$]  & $<-2.92$ & $<-2.23$ & $<-4.36$ & $<-4.41$ & - & $<-4.13$ & $<-3.85$ & $<-4.23$ & $<-3.17$ & $<-2.96$ & -\\
     $\rm \log SO_2$ & $\mathcal{U}$[$-12$, $-1$]  & $-5.67^{+0.31}_{-0.32}$ & $-5.63^{+0.60}_{-0.59}$ & $-6.01^{+0.30}_{-0.39}$ & $-6.02^{+0.31}_{-0.41}$ & - & $-5.47^{+0.29}_{-0.28}$ & $-6.63^{+0.94}_{-3.09}$ & $-5.51^{+0.32}_{-0.35}$ & $-6.35^{+0.38}_{-1.01}$ & $-5.52^{+0.29}_{-0.28}$ & - \\
     $\rm \log H_2S$ & $\mathcal{U}$[$-12$, $-1$]  & $<-3.99$ & $<-3.49$ & $<-4.18$ & $<-4.21$ & - & - & - & - & - & - & - \\
     $\rm \log Na$ & $\mathcal{U}$[$-12$, $-1$]  & $-4.10^{+0.47}_{-0.52}$ & $-4.29^{+1.00}_{-1.29}$ & $-3.59^{+0.47}_{-0.55}$ & $-3.53^{+0.45}_{-0.48}$ & - & $-3.99^{+0.64}_{-0.81}$ & $<-2.18$ & $<-2.37$ & $<-3.16$ & $-4.15^{+0.63}_{-0.76}$ & - \\
     $\rm \log K$ & $\mathcal{U}$[$-12$, $-1$]  & $<-5.45$ & $<-4.57$ & $<-5.01$ & $<-5.12$ & - & $<-4.89$ & $<-3.68$ & $<-4.00$ & $<-5.35$ & $<-4.92$ & - \\
     $\rm \log NH_3$ & $\mathcal{U}$[$-12$, $-1$]  & $<-5.47$ & $<-3.91$ & $<-5.87$ & $<-5.89$ & - & $<-5.67$ & $<-4.97$ & $<-5.49$ & $<-5.46$ & $<-5.49$ & - \\
     $\rm \log HCN$ & $\mathcal{U}$[$-12$, $-1$] & $<-4.64$ & $<-3.68$ & $<-5.48$ & $<-5.46$ & - & $<-4.41$ & $<-4.59$ & $<-5.05$ & $-5.67^{+0.34}_{-0.36}$ & $<-4.67$ & - \\ 
     $\rm \log P_{cloud}$ [bar] & $\mathcal{U}$[$-6$, 2] & $>-1.91$ & $>-2.92$ & $>-1.86$ & $>-1.77$ & $>-2.36$ & $>-2.07$ & $>-2.71$ & $>-2.14$ & $>-1.49$ & $>-1.90$ & - \\
     $\rm f_{Ray}$$^{e}$ & $\mathcal{U}$[0, 10] & $1.51^{+0.34}_{-0.31}$ & $0.58^{+0.74}_{-0.69}$ & $1.80^{+0.27}_{-0.27}$ & $2.67^{+0.23}_{-0.22}$ & $2.46^{+0.34}_{-0.32}$ & $1.15^{+0.26}_{-0.24}$ & $2.03^{+0.43}_{-0.40}$ & $2.04^{+0.36}_{-0.37}$ & $1.57^{+0.28}_{-0.25}$ & $1.31^{+0.34}_{-0.28}$ & $2.88^{+0.60}_{-0.53}$ \\
     $\rm \alpha_{Ray}$$^{e}$ & $\mathcal{U}$[$-5$, 5] & $-2.28^{+0.36}_{-0.41}$ & $-2.29^{+0.71}_{-0.72}$ & $-2.11^{+0.32}_{-0.31}$ & $-1.97^{+0.31}_{-0.28}$ & $-3.36^{+0.18}_{-0.18}$ & $-1.51^{+0.23}_{-0.24}$ & $-2.34^{+0.38}_{-0.44}$ & $-2.64^{+0.39}_{-0.47}$ & $-2.96^{+0.34}_{-0.37}$ & $-1.89^{+0.29}_{-0.32}$ & $-3.22^{+0.77}_{-0.64}$\\[2.5pt]
     $\rm [M/H]$ & $\mathcal{U}$[$-2$, 3] & - & - & - & - & $1.14^{+0.15}_{-0.15}$ & - & - & - & - & - & $0.81^{+0.13}_{-0.15}$ \\
     $\rm C/O$ & $\mathcal{U}$[0, 1.5] & - & - & - & - & $0.25^{+0.13}_{-0.09}$ & - & - & - & - & - & $0.44^{+0.10}_{-0.09}$ \\[2.5pt]
    \enddata
    \tablecomments{Results reported as median and 1-$\sigma$ range for bounded posteriors and 3-$\sigma$ limits otherwise. - indicates that a parameter was not included in a retrieval.\\
     $^{a}$ Retrievals performed on combined VLT + \texttt{exoTEDRF} JWST spectrum, with JWST data binned to $R=300$.
     $^{b}$ Same as $a$, but using the alternate \texttt{NAMELESS} NIRISS and \texttt{Eureka!}\ NIRSpec JWST spectra. 
     $^{c}$ Same as $a$, but including Magellan/IMACS and HST/WFC3 data. 
     $^{d}$ Retrievals performed on \texttt{exoTEDRF} JWST data only.
     $^{e}$ Isothermal atmosphere temperature for \texttt{POSEIDON}, and top-of-atmosphere temperature for \texttt{NemesisPy}, \texttt{Pyrat Bay}, and \texttt{Aurora}, following the PT parameterization of \citet{madhusudhan_temperature_2009}.
     $^{f}$ $f\rm _{Ray}$ is the Rayleigh enhancement factor and $\rm \alpha_{Ray}$ the scattering slope for ``haze scattering" opacity.
                  }
\end{deluxetable*}
\end{rotatetable*}

\subsection{Atmospheric Composition of WASP-96 b}
\label{sec: Atmosphere Composition}

\begin{figure}
    \centering
    \includegraphics[width=\columnwidth]{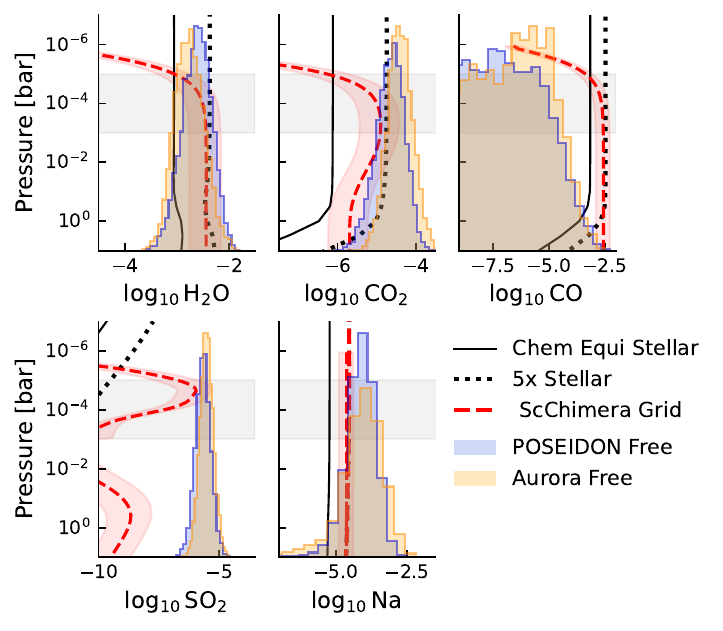}
    \caption{Abundances of several prominent chemical species inferred from the \texttt{POSEIDON} (blue histograms) and \texttt{Aurora} (orange histograms) free retrievals. Overplotted are chemical equilibrium abundance profiles for a stellar (logZ$\sim$2$\times$ solar, C/O=0.42; solid) and 5$\times$ stellar (logZ$\sim$10$\times$ solar; dotted) metallicity atmosphere, as well as constraints from the self-consistent grid (red). The grey shaded regions denote the approximate pressures probed by these observations. Solar abundances are taken to be those from \citet{asplund2009} to match \citet{nikolov_solar--supersolar_2022}.}
    \label{fig:Abundance_profiles}
\end{figure}

Particularly with the addition of our new NIRSpec/G395H data, we reveal a rich transmission spectrum for WASP-96\,b. In Figure~\ref{fig:Abundance_profiles}, we compare the retrieved abundances of several key species from the \texttt{POSEIDON} and \texttt{Aurora} free retrievals to predictions from chemical equilibrium and findings from the \texttt{ScCHIMERA} self-consistent grid. Prominent H$_2$O and CO$_2$ features are visible and we retrieve abundances spanning log\,VMR $-2.82^{+0.62}_{-0.76}$ to $-2.30^{+0.38}_{-0.41}$ and $-4.91^{+0.30}_{-0.30}$ to $-4.15^{+0.70}_{-0.77}$ respectively depending on the data and retrieval setup. The H$_2$O abundance is consistent with the findings of \citet{nikolov_solar--supersolar_2022} and \citet{wang_comprehensive_2026}, but slightly higher than that of \citet{taylor_awesome_2023}. 

Na is also clearly visible in the spectrum though its abundance is only constrainable with the contributions of the VLT data --- JWST observations alone yield an upper limit \citep[e.g.,][]{taylor_awesome_2023}. There is more variance in the retrieved Na abundance compared to other species, with values ranging from log\,VMR $-4.29^{+1.00}_{-1.29}$ to $-3.53^{+0.45}_{-0.48}$ depending on the data and retrieval setup. These constraints are consistent with \citet{wang_comprehensive_2026} and slightly more elevated than found by \citet{nikolov_solar--supersolar_2022} (though still consistent at $\lesssim$2$\sigma$ in all cases). 

We find moderate evidence for the presence of SO$_2$ with abundances ranging from log\,VMR $-6.01^{+0.30}_{-0.39}$ to $-5.47^{+0.29}_{-0.28}$ across the different retrievals. We discount the \texttt{Pyrat Bay} R=100 retrieval here since the SO$_2$ feature is not fully resolved at this resolution and its abundance is not well constrained. To quantify the model preference for the inclusion of SO$_2$, we ran an additional \texttt{NemesisPy} retrieval using the nominal setup but leaving out SO$_2$. We find an evidence value of $\ln Z$=$-$596.57 compared to $-$593.88 for the nominal setup, yielding a Bayes factor of $\ln B$=2.69 --- or moderate evidence for SO$_2$ using the Jeffreys scale \citep{Jeffreys_1935}.

As shown in Figure~\ref{fig:Abundance_profiles} the \texttt{ScCHIMERA} + \texttt{Photochem} SO$_2$ abundance is well-matched to the values from the free-retrievals at the pressure levels probed by our observations. Based on its equilibrium temperature and atmospheric metallicity, WASP-96\,b falls directly on the SO$_2$ shoreline derived by \citet{crossfield_mapping_2025}, and our retrieved SO$_2$ abundance agrees with their grid, assuming a 10--20$\times$ solar atmosphere metal enrichment. If confirmed via follow-up observations (e.g., MIRI/LRS) this finding would validate the \citet{crossfield_mapping_2025} shoreline in a higher-temperature regime (T$\rm _{eq}$$\sim$1300\,K compared to e.g., $\sim$1100\,K for WASP-39\,b).

On the other hand, we do not find any evidence for the presence of CO, which should be a major carrier of both O and C at these temperatures \citep{moses_disequilibrium_2011, madhusudhan_co_2012}. Our inability to place a bounded constraint on CO is likely driven by a combination of insufficient precision redwards of 4.5\,µm where CO is most visible and the presence of a strong CO$_2$ feature at 4.3\,µm which can hide part of the CO absorption band in transmission spectra. Indeed, with the exception of WASP-107\,b and WASP-39\,b, other planets with robust CO detections in transit have had weak or non-existent CO$_2$ features \citep{kirk_bowie-align_2025, ahrer_escaping_2025, meech_bowie-align_2025}. WASP-107\,b, though, is a super-puff with massive atmosphere features which allows the easy identification of molecular features \citep[e.g.,][]{welbanks_high_2024}, and in WASP-39\,b CO was identified through a cross-correlation analysis instead of the standard retrieval approach.

\begin{figure*}
    \centering
    \includegraphics[width=0.95\linewidth]{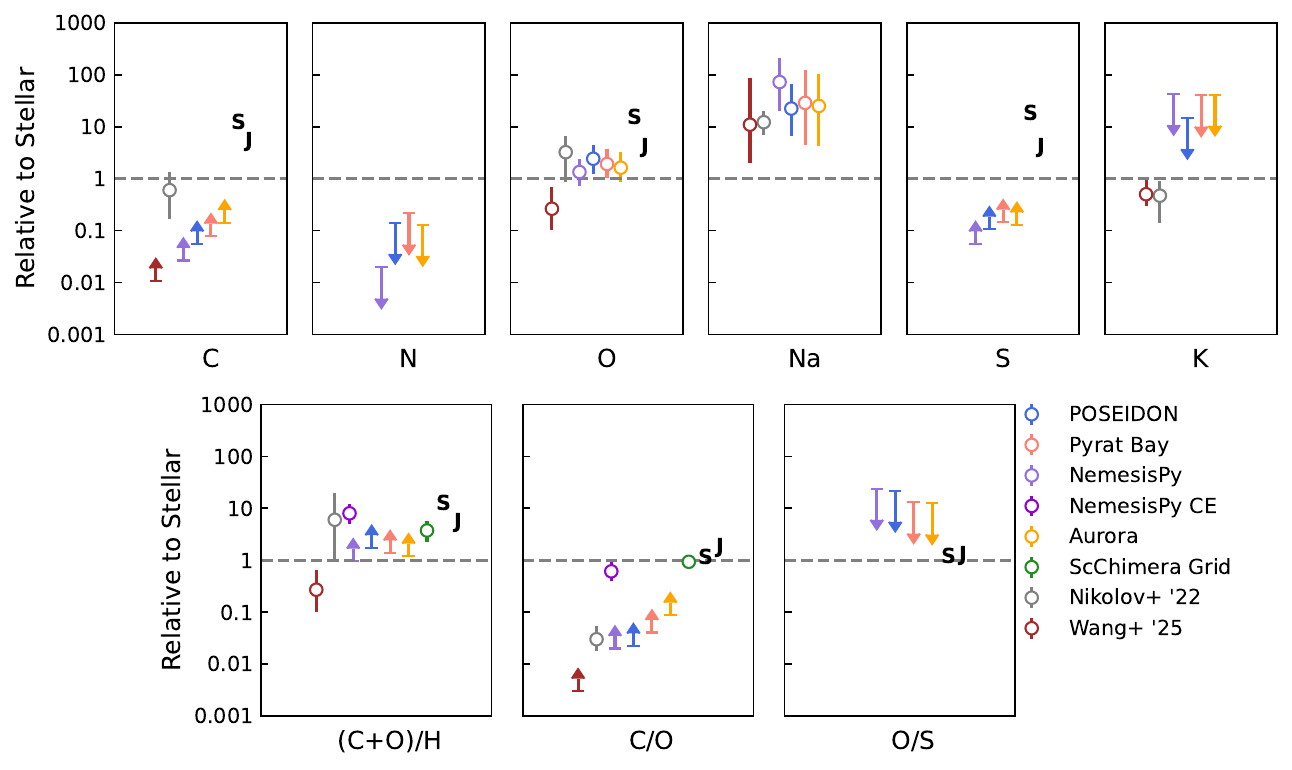}
    \caption{Elemental abundances and ratios in the atmosphere of WASP-96\,b.
    \emph{Top row}: Derived elemental abundances for WASP-96\,b's atmosphere from each retrieval code (coloured points) normalized to stellar values \citep{nikolov_solar--supersolar_2022}. For bounded constraints, the posterior median and 1-$\sigma$ confidence interval are shown. Otherwise, 3-$\sigma$ upper limits are plotted. The approximate locations of Jupiter and Saturn (sourced from \citealp{atreya_origin_2024}) are marked with bold letters. 
    \emph{Bottom row}: Abundance ratios derived from the above elemental constraints.}
    \label{fig:Abundances_to_Star}
\end{figure*}

In general, with the exception of CO our freely-retrieved abundances agree well with chemical equilibrium predictions and the results from the self-consistent grid (Figure~\ref{fig:Abundance_profiles}). Additionally, Figure~\ref{fig:Abundances_to_Star} summarizes our findings for both elemental abundances and ratios, normalized to stellar values (from \citealp{nikolov_solar--supersolar_2022}). Taken together, the freely-retrieved H$_2$O and CO$_2$ abundances indicate metallicity (as C+O/H) lower limits ranging from 1--4$\times$ stellar or 2--8$\times$ solar, depending on the particular retrieval code. The lack of constraint on the abundance of CO, which should be a prominent carrier of C in particular, prevent a bounded metallicity constraint from free retrievals alone. However, with CO self-consistently included, the \texttt{ScCHIMERA} grid finds a metallicity of 2--6$\times$ stellar (4--12$\times$ solar). The \texttt{NemesisPy} chemical equilibrium retrievals yield an even higher metallicity of 5--12$\times$ stellar ($\sim$10--20$\times$ solar), though, of all models it provides the worst fit to the spectrum, not capturing the SO$_2$ feature in particular. This is due to \texttt{FastChem} only considering thermochemical equilibrium processes, whereas SO$_2$ is photochemically produced --- further underlining the importance of photochemistry in WASP-96\,b's atmosphere. In general, though, all models indicate a solar-to-super-solar metallicity atmosphere heavily influenced by photochemistry. These results also highlight the importance of considering multiple modelling schemes with varying amounts of flexibility, since critical abundances and ratios can be difficult to measure with free retrievals alone. 

We obtain bounded constraints on CH$_4$ only in our \texttt{Pyrat Bay} retrievals (which use different, higher-resolution opacities; see Section~\ref{sec: Pyrat Bay} and Table~\ref{tab: Opacities}), at abundances far above chemical equilibrium expectations for the upper atmosphere. As demonstrated by \citet{zamyatina_quenching-driven_2024}, due to longitudinal quenching, CH$_4$ could be present in detectable abundances in WASP-96\,b's terminator atmosphere at lower metallicities (1$\times$ solar), but not at higher values (10$\times$ solar). Moreover, strong vertical mixing could also cause CH$_4$ to quench in the deep atmosphere where its abundance is more comparable to the values found by \texttt{Pyrat Bay}. However, even if substantial quantities of CH$_4$ are transported to the terminator upper atmosphere, it is unlikely that they would be able to persist due to the susceptibility of CH$_4$ to photodissociation \citep{Fleury_2023}. Finally, the retrievals also do not reveal evidence for H$_2$S, K, HCN, or NH$_3$.

Our inferred composition is in broad agreement with previous studies of the atmosphere of WASP-96\,b \citep[e.g.,][]{nikolov_solar--supersolar_2022, mcgruder_access_2022, yip_compatibility_2021}. As shown in Figure~\ref{fig:Abundances_to_Star}, our O and Na abundances are consistent with \citet{nikolov_solar--supersolar_2022}. However, we do not find evidence for K whereas \citet{nikolov_solar--supersolar_2022} do. We are, though, able to place more robust constraints on the atmosphere metallicity as the presence of CO$_2$ implies values above solar. On the other hand, our results are in tension with the findings of \citet{wang_comprehensive_2026}, particularly their inference of a sub-stellar metallicity which is incompatible with our robust detection of CO$_2$. This highlights the need to simultaneously analyze multiple chemical species to place robust constraints on atmosphere metallicity.

\subsubsection{Implications for Planet Formation}
\label{sec: Formation}

Planet atmospheres are predicted to retain clues into the planet's formation and migration history \citep{oberg_effects_2011, mordasini_imprint_2016, crossfield_volatile--sulfur_2023, chachan_breaking_2023, kirk_bowie-align_2025}. Elemental ratios, in particular, are key in this endeavour \citep[e.g.,][]{oberg_effects_2011, chachan_breaking_2023} and since we have evidence for O-, C-, and S-bearing species, we can begin to construct multiple such ratios. The composition of WASP-96 itself is also well characterized \citep{nikolov_solar--supersolar_2022} meaning we can compare elemental ratios derived from the atmosphere of WASP-96\,b to that of the host star, which are widely used as representative of the proto-stellar environment.

As shown in the lower panels of Figure~\ref{fig:Abundances_to_Star}, the free retrievals yield a lower limit on the C/O ratio --- enormously influenced by the non-detection of CO, which should be the dominant carried of C in WASP-96\,b's atmosphere. With the inclusion of additional physics constraints, the chemical equilibrium and self-consistent grid models obtain bounded constraints of $0.25^{+0.13}_{-0.09}$ and $0.44^{+0.10}_{-0.09}$ respectively --- sub-stellar-to-stellar given the C/O ratio of WASP-96 is 0.42 \citep{nikolov_solar--supersolar_2022}.

WASP-96\,b's broadly super-stellar metallicity and stellar-to-super-stellar alkali abundances matches the results from \citet{welbanks_massmetallicity_2019}. Considering the sub-stellar-to-stellar derived C/O ratios, our result joins similar findings for giant transiting planets \citep[e.g.,][]{ahrer_early_2023, meech_bowie-align_2025, kirk_bowie-align_2025}. In combination, these two factors generally indicate formation beyond the H$_2$O snowline, followed by disk-driven migration and the accretion of O-rich solid material, which can elevate the planet's metallicity and lower its C/O ratio \citep{espinoza_metal_2017, ali-dib_disentangling_2017, cridland_connecting_2019, penzlin_bowie-align_2024}.

Recent theoretical work has demonstrated that moving beyond the standard metallicity vs.\ C/O prescriptions and considering e.g., refractory-to-volatile ratios can break degeneracies in planet formation scenarios \citep[e.g.,][]{lothringer_new_2021, chachan_breaking_2023, crossfield_volatile--sulfur_2023}. In particular, \citet{crossfield_volatile--sulfur_2023} suggested using the S abundance as a tracer of a planet's refractory content, thereby allowing for the construction of refractory-to-volatile ratios with NIR spectra alone. However, without a constraint on the abundance of H$_2$S, which should be the dominant S carrier at the temperatures of WASP-96\,b's atmosphere \citep{tsai_photochemically_2023}, we can only derive a relatively unconstraining O/S ratio, compatible with multiple formation scenarios. This highlights the importance of H$_2$S to the goal of understanding S chemistry in exoplanet atmospheres. Unfortunately, the main NIR absorption feature from H$_2$S at $\sim$3.9\,µm falls in the NIRSpec/G395H detector gap. Future observations should consider using the G395M grating (or PRISM if possible) which does not have a detector gap, if a primary goal is to constrain a planet's S inventory.

\subsection{Aerosols After All?}
\label{sec: Aerosols}

Across all model tests and data combinations we find that aerosol-free models cannot adequately fit the data. When using the cloud-haze parameterization the opaque cloud deck is generally placed below the observable photosphere, but a scattering slope is consistently inferred. If due to aerosols, sloped opacity requires the presence of small particles \citep{wakeford_transmission_2015}, irrespective of whether they are photochemically produced (i.e., hazes) or condensates (i.e., clouds). Both sources are possible \textit{a-priori}, with the temperature of WASP-96\,b being ideal for cloud formation (e.g., \citealp{samra_clouds_2023}; Figure~\ref{fig:GCM_PT_Profiles}) and the presence of SO$_2$ indicating the effectiveness of photochemistry on WASP-96\,b.

Our results stand in contrast to most previous observational studies of WASP-96\,b which have not seen the same optical slope as we do here \citep[e.g.,][]{nikolov_absolute_2018, yip_compatibility_2021, nikolov_solar--supersolar_2022, mcgruder_access_2022}. These previous analyses, though, required offsets between ground-based and HST spectra, which without overlapping wavelength coverage can be difficult to constrain \citep{yip_compatibility_2021, mcgruder_access_2022, nikolov_solar--supersolar_2022}. As demonstrated by \citet{radica_awesome_2023}, an offset between the VLT and HST spectrum could mask the slope seen here. 

We are though, in line with the findings of \citet{radica_awesome_2023} and \citet{taylor_awesome_2023} who analyzed only the JWST NIRISS data. Those studies also highlight the additional degeneracy between the abundance of Na and the presence of a scattering slope, particularly since NIRISS only covers the red wing of the Na feature. Due to our addition of ground-based optical data which fully resolves the Na feature, this degeneracy is more mitigated. However, we cannot fully discount the possibility that the slope is caused by broad Na wings and that our current opacities are insufficient to model it.

Alternatively, the slope in our data could be attributed to the effects of stellar contamination via the transit light source effect \citep[TLSE;][]{rackham_transit_2018, rackham_transit_2019}. We explore this possibility with \texttt{Aurora}, this time adding a stellar contamination parameterization \citep[following e.g.,][]{fournier-tondreau_near-infrared_2024} to the baseline model. In this case, we find that a star with $\sim$5\% spot coverage (at $\rm T_{spot}\approx3900$\,K) can provide a slope across the VLT and JWST bandpass that can mimic the scattering effects of an aerosol layer. In this case, the effects of the aerosol layer are significantly weaker and consistent with Rayleigh-like scattering in the atmosphere of WASP-96\,b. Additionally, the inferred stellar parameters are consistent with expectations for a late G-type star like WASP-96 \citep{rackham_transit_2019}, making this another plausible explanation for the transmission spectrum slope. 


On the other hand, when \citet{wang_comprehensive_2026} jointly analyzed a re-reduction of the NIRISS/SOSS observations with the VLT transit data, they found a NIRISS spectrum significantly flatter than those produced in this work (or used in \citealp{radica_awesome_2023} and \citealp{taylor_awesome_2023}). They interpreted the data to show evidence for aerosols, but in the form of grey cloud opacity rather than a scattering slope, since their spectrum did not display an optical slope.  They note that this difference may be due to the use of newer calibration reference files than were available to earlier studies. However, here we have also made use of the most up-do-date calibration files available from the CRDS, and so this is unlikely to be the true driver of the differences between the two spectra. We note, however, that although \citet{wang_comprehensive_2026} carry out a group-level 1/$f$ correction in their re-analysis, they do not do a group-level background subtraction, which, as shown by \citet{radica_awesome_2023}, can cause substantial wavelength-dependent biases. Determining whether that is the ultimate cause of the differences between our analysis and theirs is beyond the scope of this work.

\section{Exploration of Morning-Evening Limb Asymmetry}
\label{sec: Limb Asymmetry}

WASP-96\,b should be a strong candidate for observable morning-evening limb asymmetry \citep{samra_clouds_2023, murphy_analytic_2024, zamyatina_quenching-driven_2024}. \citet{zamyatina_quenching-driven_2024}, in particular, explored the observability of limb asymmetry on WASP-96\,b using the Met Office's \texttt{Unified Model} (UM) GCM, finding the potential for up to 500\,ppm differences between the morning and evening limbs, depending on the assumed metallicity and whether or not the planet's atmosphere is in chemical equilibrium. 

\begin{figure}
    \centering
    \includegraphics[width=0.9\columnwidth]{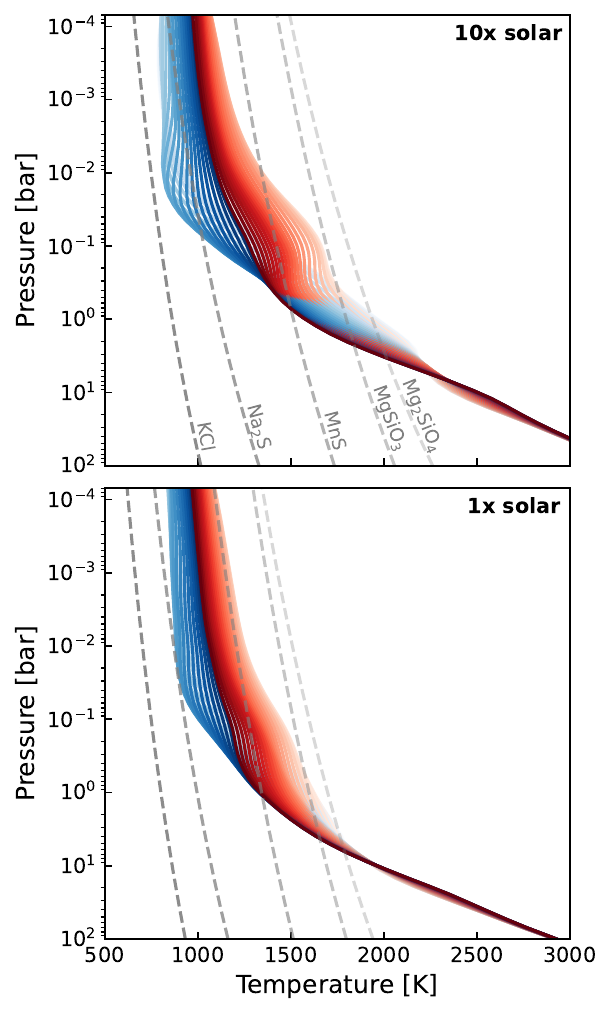}
    \caption{PT profiles derived from cloud-free UM GCM simulations of WASP-96\,b at solar (\emph{bottom}) and 10$\times$ solar (\emph{top}) metallicity, which bracket our derived atmosphere composition. Blue, and red profiles represent the morning and evening hemispheres, respectively. Levels of fading denote latitude, with the boldest colours being polar latitudes. Condensation curves for prominent condensate species (labelled) are shown with grey dashed lines.}
    \label{fig:GCM_PT_Profiles}
\end{figure}

From Figure~\ref{fig:Limb_Spectra}, the morning and evening spectra show hints of asymmetry. Though very low S/N, the morning spectrum appears flatter across the full waveband explored, whereas the evening spectrum show suggestions of features, particularly around 1.4 and 4.5\,µm where H$_2$O and CO$_2$ absorb. We also show the T$_0$ spectrum in the bottom panel of Figure~\ref{fig:Limb_Spectra}, derived from spectroscopic fits to the transit light curves assuming a uniform planet (i.e., a \texttt{batman} model) and allowing the mid-transit time to vary. We see evidence for the same mid-transit time offset in the Na feature as \citet{wang_comprehensive_2026}, though, this is the lowest-S/N part of the data, and thus potentially the least reliable region to constrain limb asymmetry.

Inspired by recent studies finding evidence for clear evening and cloudy morning limbs on giant exoplanets \citep[e.g.,][]{murphy_panchromatic_2025, mukherjee_cloudy_2025, fu_overcast_2025} we explore the possibility for WASP-96\,b to have asymmetric cloud coverage on its limbs using the UM GCM. Specific details of the GCM setup can be found in Appendix~\ref{app: GCM}, and we show generated PT profiles in Figure~\ref{fig:GCM_PT_Profiles} at two representative metallicities. We also overplot condensation curves from \citet{morley_neglected_2012} and \citet{visscher_atmospheric_2010} for some prominent condensate species. The UM PT profiles support the condensation of species like Na$_2$S at $\sim$mbar levels on the cooler morning terminator, whereas the comparatively warmer evening terminator would remain relatively cloud free. This is in rough agreement with the findings of \citet{samra_clouds_2023} who concluded that mild asymmetry is possible in WASP-96\,b's atmosphere for a variety of condensate compositions. We also overplot morning and evening limb-averaged spectra from the aerosol-free 10$\times$ solar UM GCM run in purple in Figure~\ref{fig:Limb_Spectra}. The model provides an excellent match to the evening spectrum, but overpredicts feature sizes in the morning. 

\subsection{Towards a More Robust Quantification of Transit Limb Asymmetry}
\label{sec: Limb Holistic}

\begin{figure}
    \centering
    \includegraphics[width=0.8\columnwidth]{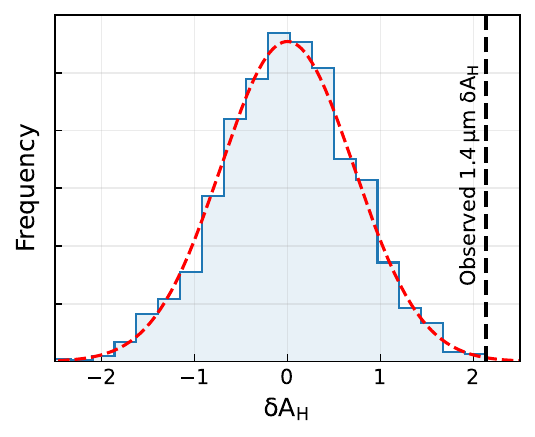}
    \caption{Distribution of $\rm \delta A_H$ values from 10000 injection-recovery tests with no underlying limb asymmetry. The red dashed curve shows the best-fitting Gaussian distribution ($\mu=-0.012, \sigma=0.717$). The observed $\rm \delta A_H$ (black dashed line) for the 1.4\,µm H$_2$O band of 2.13 is in the 99.72$^{\rm th}$ percentile of the distribution, making it a $\sim$2.6-$\sigma$ outlier. Read another way, potential morning-evening asymmetry around 1.4\,µm is $\sim$2.6-$\sigma$ significant based on this test.}
    \label{fig:Inject_Recovery_AH}
\end{figure}

The above discussion, though, is only circumstantial evidence for limb asymmetry in WASP-96\,b. The question arises of how to robustly quantify the presence of limb asymmetry, particularly in low-S/N cases like the present one. The T$_0$ spectrum is an excellent tracer of limb asymmetry as variation in the mid-transit time can only be caused by changes in shape of the planet when all else is held fixed \citep{seager_unique_2003, espinoza_constraining_2021, murphy_analytic_2024}. But, these spectra can also be low S/N (as in the present case), and only really yield ``by-eye" evidence. On the opposite end of the spectrum, Bayesian model comparison tests have been performed on light curves in order to determine whether there is sufficient evidence for an asymmetric \texttt{catwoman} model over a symmetric \texttt{batman} one \citep[e.g.,][]{ahrer_tracing_2025}. As in \citet{ahrer_tracing_2025}, we find that the symmetric model is strongly favoured ($\ln B \gtrsim 5$) for WASP-96\,b at all wavelengths. However, this test is not ideal as it does not take into account wavelength-dependent information (i.e., the presence and lack of spectral features). 

Here, we suggest some tests to quantify the degree to which morning and evening spectra show feature differences. We first perform a Gaussian feature test \citep[e.g.,][]{moran_high_2023, may_double_2023, taylor_jwst_2025} on the morning and evening spectra. Following the methods of \citet{taylor_jwst_2025}, we compare two models: a flat line (one free parameter; an offset) and a model with a single Gaussian feature (four free parameters; an offset, Gaussian feature position, amplitude, and width). We treat the NIRISS and NIRSpec separately on account of the potential for an offset between them. Using \texttt{dynesty} \citep{speagle_dynesty_2020}, we fit each model to the morning and evening spectrum from each instrument and calculate the associated Bayesian evidence values. 

For the morning spectra, the flat line model is strongly preferred ($\ln B > 3.5$) for both instruments over the presence of a Gaussian feature, supporting the claim that both are featureless. On the other hand, the test finds weak-to-moderate evidence ($\ln B = 1.36$) for a Gaussian bump at 4.5\,µm in the NIRSpec evening spectrum compared to a flat line. The NIRISS evening spectrum shows marginal evidence for a Gaussian feature at $\sim$1.4\,µm ($\ln B = 0.76$) over a flat line, though this increases to $\ln B = 1.21$ if we force the feature to be at 1.4\,µm instead of allowing the prior to span the whole SOSS waveband. This test, though, is also not entirely foolproof. It works if one limb is entirely featureless whereas the other shows features. However, it would not be informative when features are present in both limbs. 

Next, we try another test to address these shortcomings which builds off of the work of \citet{fu_overcast_2025}. We define a spectral amplitude index, $A_H$, which quantifies the difference in transit depth inside and outside of a spectral feature:
\begin{equation}
    A_H = (\textrm{in-band} - \textrm{out-of-band})/\textrm{error}.
\end{equation}
We focus on the NIRISS 1.4\,µm H$_2$O band (in-band: 1.35--1.5\,µm, out-of-band: 0.9--1.3\,µm, shaded blue and grey respectively in Figure~\ref{fig:Limb_Spectra}) the same as in \citet{fu_overcast_2025}. The in- and out-of-band values are band-averages in atmosphere scale heights, and the errors are propagated from the individual transit depths and added in quadrature \citep{fu_overcast_2025}. We calculate $A_H$ separately for the morning and evening limbs and then subtract them to quantify the difference in the H$_2$O-band amplitude,
\begin{equation}
    \delta A_H = A_{H, \mathrm{evening}} - A_{H, \mathrm{morning}}.
\end{equation}
The value for WASP-96\,b is $\delta A_H$=2.13$\pm$0.31.

Next, we compare to the distribution of $\delta A_H$ values in the absence of any underlying asymmetry. To this end, we simulate 10000 sets of light curves with the same noise and transit properties as WASP-96\,b, except that the planet is uniform (i.e., no limb asymmetry). We then fit these light curves with \texttt{catwoman} to obtain limb spectra and calculate $\delta A_H$ for each case as was done above. The calculated distribution of $\delta A_H$ values are shown in Figure~\ref{fig:Inject_Recovery_AH}, and is well-described by a Gaussian centered on zero, with a width of 0.72. This analysis suggests that differences of this size can arise stochastically at low S/N. However, asymmetry in the 1.4\,µm band at the level we see is a 2.6-$\sigma$ outlier in the derived distribution. Read another way, the potential 1.4\,µm feature asymmetry is 2.6-$\sigma$ significant via this test. We also explore using the 4.5\,µm CO$_2$ feature for this test, and find similar but weaker results, primarily due to the lower overall S/N in this region. 

Again though, this test is not entirely foolproof. For example, it would be inappropriate in cases where limb differences are solely due to temperature or where suitable out-of-band wavelengths cannot be identified. We suggest that the development of a robust and widely-applicable quantification scheme for limb asymmetry should be a community priority. For example, ``leave-one-out" cross-validation techniques \citep[e.g.,][]{cloutier_confirmation_2019, radica_revisiting_2022} recently adapted for the interpretation of atmospheric models and inferences \citep{welbanks_application_2023, nixon_methods_2024} could potentially be applied to the interpretation of limb-asymmetries, building on the work of \cite{Challener2023} for eclipse mapping.

\section{Conclusions} 
\label{sec: Conclusion}

In this work we conducted an in-depth characterization of the atmosphere of WASP-96\,b, building off of the work by \citet{radica_awesome_2023} and \citet{taylor_awesome_2023}. We combined archival VLT/FORS2 and JWST NIRISS/SOSS with new NIRSpec/G395H transit observations to create a high precision 0.35 -- 5\,µm atmosphere spectrum, and we summarize the major findings of our analysis below.

\begin{enumerate}
    \item We obtain strong detections of H$_2$O, CO$_2$, and Na in free retrievals, however CO remains unconstrained. Free retrievals thus indicate a broadly super-stellar metallicity (lower limits of 2--8$\times$ solar or 1--4$\times$ stellar depending on the particular retrieval code). With CO included self-consistently, the \texttt{ScCHIMERA} grid yields a metallicity of 4--12$\times$ solar (2--6$\times$ stellar). 
    
    \item We find a moderate ($\ln B$=2.69) preference in free retrievals for models with SO$_2$ versus those without. The retrieved abundance agrees with photochemical predictions, and WASP-96\,b falls right on the SO$_2$ shoreline proposed by \citet{crossfield_mapping_2025}.
    \item Our chemical equilibrium retrievals and self-consistent grids yield a sub-stellar-to-stellar C/O ratio. When combined with our metallicity constraints, this potentially indicates formation beyond the H$_2$O snowline and the accretion of volatile-rich material. 
    
    \item Our atmosphere spectrum displays a strong slope bluewards of $\sim$1.5\,µm which our models explain via aerosol scattering opacity. Small-particle condensate clouds or photochemically-produced hazes can potentially cause this slope, though we also cannot entirely rule out the broad wings of the Na feature, or stellar contamination. Further observations should be conducted to disentangle/uniquely identify these factors.
    
    \item We explore the possibility for limb asymmetry in the atmosphere of WASP-96\,b, but do not find conclusive evidence one way or the other. We encourage the community to prioritize the development of a metric to quantify the presence of limb asymmetry, and we present some suggestions in this direction.
    
    \item Finally, we demonstrate that removing 1/$f$ noise in JWST NIRISS/SOSS observations after ramp-fitting (i.e., at the integration-level) can inject an optical-to-NIR slope and excess covariance into transmission spectra compared to a group-level correction. 
\end{enumerate}

Our identification of SO$_2$ is not definitive and should be followed up with JWST MIRI/LRS observations which are incredibly sensitive to SO$_2$ \citep[e.g.,][]{powell_sulfur_2024,dyrek_so2_2024} as well as cloud-induced limb asymmetry \citep[e.g.,][]{murphy_panchromatic_2025}. Moreover, MIRI observations would provide definitive evidence for aerosols via the identification of specific absorption features \citep{wakeford_transmission_2015, grant_jwst-tst_2023}. Finally, we reiterate our encouragement that the community to be thoughtful and thorough when reporting limb asymmetry detections, or lack thereof --- particularly in low-S/N regimes.

\begin{acknowledgments}
M.R.\ would like to acknowledge funding from the Natural Sciences and Engineering Research Council of Canada (NSERC), as well as the Canadian Space Agency (CSA).
J.T.\ was supported by the Glasstone Benefaction, University of Oxford (Violette and Samuel Glasstone Research Fellowships in Science 2024). 
D.C.\ receives funding from the Max Planck Society.
N.J.M.\ and M.Z.\ acknowledge support from a UKRI Future Leaders Fellowship [Grant MR/T040866/1], a Science and Technology Facilities Funding Council Nucleus Award [Grant ST/T000082/1], and the Leverhulme Trust through a research project grant [RPG-2020-82]. 
J.B.\ acknowledges the support received in part from the NYUAD IT High Performance Computing resources, services, and staff expertise.
This work is based on observations made with the NASA/ESA/CSA JWST.  The data were obtained from the Mikulski Archive for Space Telescopes at the Space Telescope Science Institute, which is operated by the Association of Universities for Research in Astronomy, Inc., under NASA contract NAS 5-03127 for JWST. The specific observations analyzed can be accessed via\,\dataset[10.17909/g5qz-cv38]{10.17909/g5qz-cv38}. This research has made use of the NASA Exoplanet Archive, which is operated by the California Institute of Technology, under contract with the National Aeronautics and Space Administration under the Exoplanet Exploration Program. 
\end{acknowledgments}

\vspace{5mm}
\facilities{JWST(NIRISS), JWST(NIRSpec), Exoplanet Archive \citep{christiansen_nasa_2025}}

\software{\texttt{astropy} \citep{astropy:2013, astropy:2018}, 
\texttt{batman} \citep{kreidberg_batman_2015},
\texttt{catwoman} \citep{jones_catwoman_2020, espinoza_constraining_2021},
\texttt{dynesty} \citep{speagle_dynesty_2020},
\texttt{emcee} \citep{foreman-mackey_emcee_2013},
\texttt{ipython} \citep{PER-GRA:2007},
\texttt{jwst} \citep{bushouse_2023},
\texttt{matplotlib} \citep{Hunter:2007},
\texttt{numpy} \citep{harris2020array},
\texttt{pymultinest} \citep{buchner_statistical_2016},
\texttt{scipy} \citep{2020SciPy-NMeth}
}

\appendix

\section{Additional Data Reductions}
\label{app: additional reductions}

\subsection{NIRISS/SOSS: \texttt{NAMELESS}}

\begin{figure*}
    \centering
    \includegraphics[width=0.75\linewidth]{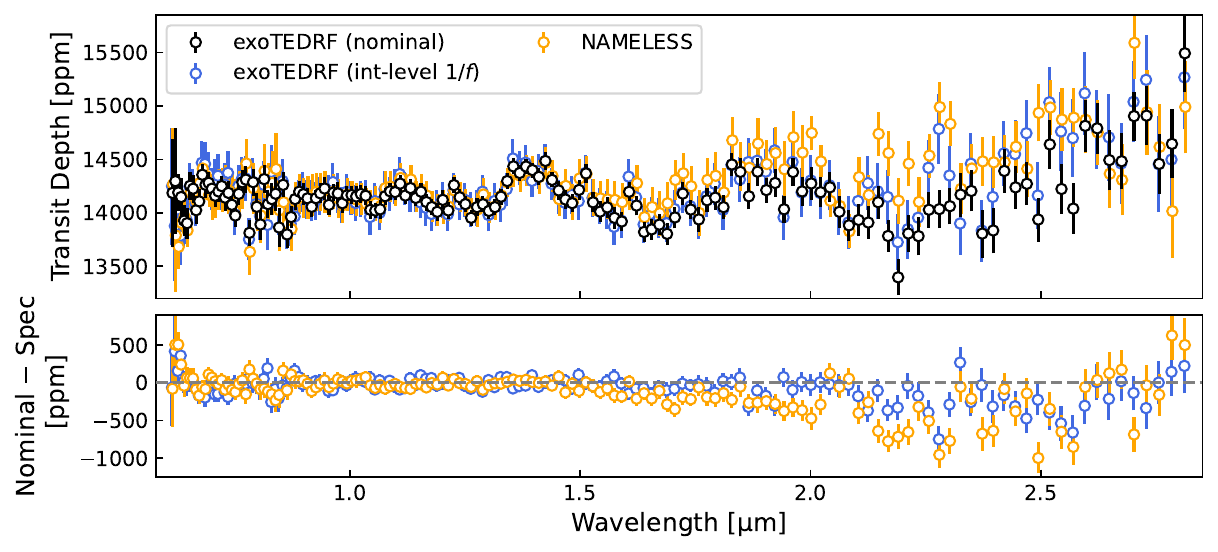}
    \caption{The differences between the \texttt{exoTEDRF} and \texttt{NAMELESS} NIRISS/SOSS spectra redwards of $\sim$1.75\,µm can be attributed to the 1/$f$ noise correction being performed at the integration  vs.\ the group level (i.e., after vs.\ before ramp-fitting). An alternate \texttt{exoTEDRF} reduction skipping the group-level 1/$f$ correction and implementing the same step at the integration level reproduces qualitatively the differences with the nominal spectrum seen with \texttt{NAMELESS}.}
    \label{fig:NIRISS_Compare_OneOverF}
\end{figure*}

We perform an independent reduction of the NIRISS/SOSS TSO using the \texttt{NAMELESS} pipeline \citep{coulombe_broadband_2023,coulombe_highly_2025}, following closely the methodology applied in \citet{coulombe_highly_2025}. We start from the raw uncalibrated data and proceed through the Stage 1 and 2 steps of the \texttt{jwst} pipeline, up to flat-field correction. We then flag bad/hot pixels by isolating pixels with large 2D second derivatives and correct for them via 2D cubic interpolation. We correct for the nonuniform background by scaling independently the two portions of the STScI model background separated by the sharp jump in flux, and subtract it from all integrations \cite{lim_atmospheric_2023}. Remaining cosmic rays are corrected for by computing the running median of all individual pixels and bringing all counts $>$4$\sigma$ to the median's value. The 1/$f$ noise is computed and subtracted by scaling each column of each trace independently and determining the 1/$f$ within the trace via $\chi^2$ minimization, following the same method outlined in \citet{coulombe_highly_2025}. Finally, we extract the stellar spectra using a fixed box width of 36 pixels.

We proceeded to fit the spectrophotometric light curves using \texttt{exoUPRF} \citep{radica_exouprf_2024, ahrer_escaping_2025, radica_promise_2025} and following the same procedure as for the \texttt{exoTEDRF} spectrum (see Section~\ref{sec: Light Curve Fits}), thereby ensuring that any differences in the transmission spectra were due to choices made during the reduction process and not introduced as part of the light curve fitting. 

\subsection{NIRSpec/G395H: \texttt{Eureka!}}

For a second independent reduction of our data we used the \texttt{Eureka!} pipeline \citep{bell_eureka_2022}, closely following the procedures used in other NIRSpec/G395H reductions of hot Jupiters \citep[e.g.,][]{kirk_bowie-align_2025,ahrer_tracing_2025}. For the first step, we used the uncalibrated fits files and followed the default steps of the \texttt{jwst} pipeline \citep{bushouse_2023} for Stage 1 and 2, with the exception of: using a $10\sigma$ threshold for the \texttt{jump\_rejection\_threshold}, using a scaling factor for the superbias to calibrate it, and subtracting a column-by-column weighted average at the group-level (masking the trace). After ramp-fitting, we ran Stage 3 which extracts the time-series spectra. We used a 5$\sigma$ threshold for spatial outlier rejection and a double-iterative $5\sigma$ threshold along the time axis. We corrected for the curvature of the spectral trace and use pixels $>8$pixels from the central trace pixels for a column-by-column weighted average background subtraction. To extract the final 1D spectra we used optimal extraction \citep{Horne1986optimalextraction} with an aperture half width of four pixels. 

We used the same wavelength bins at R=100 and R=300 as the \texttt{exoTEDRF} reduction to produce the spectroscopic light curves of the transit of WASP-96\,b. Using \texttt{Eureka!}'s Stage 4 we masked outliers in the binned light curves using a $>$5$\sigma$ threshold, a 20-pixel rolling median and five iterations.  
The extracted light curves are then fitted within Stage 5 using a \texttt{batman} transit model \citep{kreidberg_batman_2015} and a linear in time. We use the 4-parameter limb-darkening law and fix the parameters $u1, u2, u3, u4$ to the generated values using \texttt{ExoTiC-LD} \citep{Grant2024ExoTiC-LD:Coefficients} following the parameters for WASP-96 ($\rm T_{eff}$=5500K, M/H=0.14, logg=4.42; \citealp{hellier_transiting_2014}) and the \texttt{stagger} grid \citep{magic_stagger-grid_2015}. We fix the system parameters to the best-fit parameters from the joint fit, Table\,\ref{tab: WLC Parameters}. Therefore our spectroscopic light curve fits consisted of four free parameters: the transit depth, two parameters for the linear and a free parameter for error inflation. We use the MCMC python package \texttt{emcee} \citep{foreman-mackey_emcee_2013} for our light curve fits. The \texttt{Eureka!} transmission spectrum at R=300 is compared to the \texttt{exoTEDRF} one in Figure\,\ref{fig:Spectra_Compare_R300}.

\section{On the Correction of 1/{\MakeLowercase f} Noise for NIRISS/SOSS}
\label{app: SOSS 1/f}

As shown in Figure~\ref{fig:Spectra_Compare_R300}, there are significant differences between the \texttt{exoTEDRF} and \texttt{NAMELESS} NIRISS/SOSS spectra redwards of $\sim$1.7\,µm, despite the excellent agreement at bluer wavelengths. Such differences have also been observed in other studies of giant planets \citep[e.g.,][]{radica_awesome_2023, fournier-tondreau_near-infrared_2024}, but not in multiple studies of smaller planets \citep[e.g.,][]{lim_atmospheric_2023, radica_promise_2025, benneke_jwst_2024} and/or brighter stars \citep[e.g.,][]{radica_muted_2024, ahrer_escaping_2025}. We could not trace this deviation back to differences in light curve fitting, which led to a reassessment of the choices made in the data reduction itself. 

One major difference between the standard \texttt{exoTEDRF} and \texttt{NAMELESS} workflows is the correction of 1/$f$ noise at the group- vs.\ the integration-level (that is, before or after ramp-fitting). We thus perform an experiment where we redo the \texttt{exoTEDRF} data reduction, but skip the group-level 1/$f$ correction and instead perform the same step after ramp-fitting, analogous to how it is carried out with \texttt{NAMELESS}. As shown in Figure~\ref{fig:NIRISS_Compare_OneOverF} this new \texttt{exoTEDRF} spectrum is in much better agreement with \texttt{NAMELESS} and qualitatively reproduces the deviations seen with the nominal \texttt{exoTEDRF} spectrum. 

\begin{figure}
    \centering
    \includegraphics[width=0.7\linewidth]{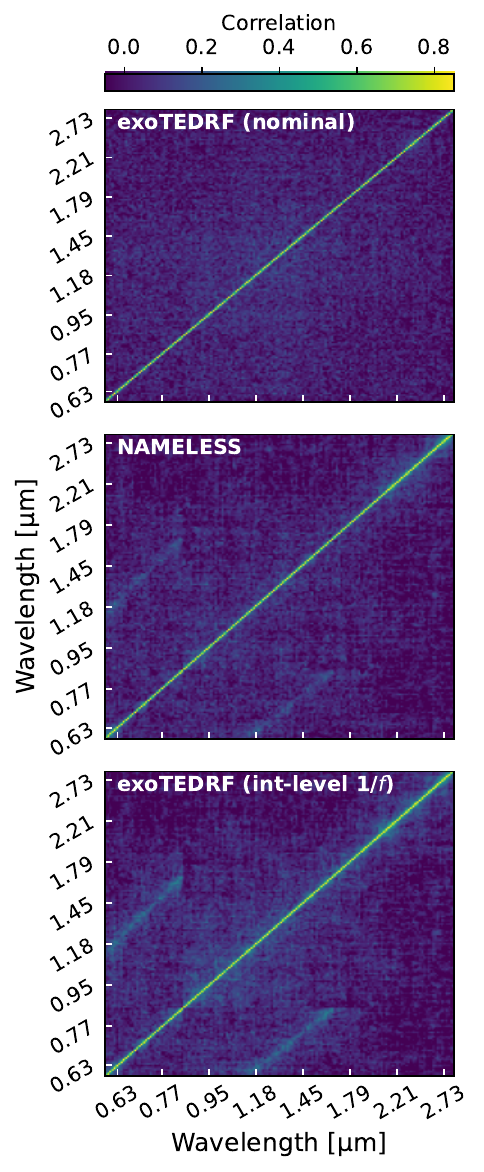}
    \caption{Covariance matrices extracted from the NIRISS/SOSS light curve fit residuals for the nominal \texttt{exoTEDRF} reduction (\emph{top}), the \texttt{NAMELESS} reduction (\emph{middle}) and the alternate \texttt{exoTEDRF} reduction performing an integration-level 1/$f$ noise correction (\emph{bottom}). Significant residual covariance between wavelengths which share a detector column are present when performing the 1/$f$ noise correction at the integration level, but not for the group-level correction.}
    \label{fig:Covariance_Matrices}
\end{figure}

Though the proximate cause seems to be the order in which one performs the 1/$f$ correction and ramp fitting, at this time the ultimate cause of these deviations is still unclear. The fact that they only appear (or potentially just appear more clearly) in datasets with deep transits, which target dimmer stars, and are largest where the instrument throughput is smallest suggest some type of diluting effect. However, despite extensive testing, we have not been able to confirm this hypothesis nor identify its root cause. 

In Figure~\ref{fig:Covariance_Matrices} we show covariance matrices for the nominal \texttt{exoTEDRF} spectrum, as well as the \texttt{NAMELESS} spectrum and the reprocessed \texttt{exoTEDRF} with integration-level 1/$f$ correction. The off-diagonal structures first noted by \citet{holmberg_exoplanet_2023} are present in the latter two, but disappear when performing the 1/$f$ correction at the group level. These off-diagonal covariances are introduced by the 1/$f$ correction, which in the three examples shown here (as well as \citet{holmberg_exoplanet_2023}) subtracts a single 1/$f$ value from each detector column, thereby introducing correlations between wavelengths in the first and second diffraction orders that share a given column. However, when 1/$f$ noise is subtracted at the group-level, the ramp-fitting erases much of these correlations yielding an almost purely diagonal, and thus ideal, covariance. 

Although we do not necessarily know the ground truth for any exoplanet transit spectrum, and it is impossible to simulate noise sources or systematics that are not well understood, we nevertheless concur with \citet{carter_effects_2025} and argue that the group-level 1/$f$ correction is the most proper method --- and is indeed how this noise source is treated for JWST's other science instruments \citep[e.g.,][]{alderson_early_2023, rustamkulov_early_2023}. Firstly, 1/$f$ noise is one of the last to be imparted on the data as it is injected during readout, and thus should be one of the first to be removed during the data reduction. The group-level 1/$f$ correction leads to lower light curve scatter and smaller final error bars \citep[e.g.,][]{radica_awesome_2023}, and finally, leaves negligible off-diagonal covariance in the data --- metrics which have been widely employed throughout the history of the field to determine the ``optimal" data treatment.

\section{Details of the UM GCM Simulations}
\label{app: GCM}

To provide an additional point of comparison for our observations, we simulate the atmosphere of WASP-96\,b using the Met Office's UM GCM which has modelled hot Jupiters and Saturns previously and been specifically used to model the atmosphere of WASP-96\,b in \citet{taylor_awesome_2023} and \citet{zamyatina_quenching-driven_2024}. The UM solves the full, deep-atmosphere Euler equations (see \citealt{wood_2014} and \citealt{mayne_2014a} for a discussion of implementation) with multi-band radiative transfer handled using the \texttt{socrates} radiative transfer code \citep{edwards_1996}.  The atmosphere is assumed to have 1$\times$ or 10$\times$ solar metallicity and with the chemical abundances initially set to be in thermochemical equilibrium.  The evolution of the abundances are followed using the UM's chemical kinetics solver using the \citet{venot_2019} C/N/O/H network, allowing for the possibility that the chemical abundances diverge from equilibrium.  A subset of these chemical species contribute opacity to the gas: H$_2$O, CO, CO$_2$, CH$_4$, NH$_3$, HCN, Li, Na, K, Rb, and Cs as well as collision-induced absorption by H$_2$-H$_2$ and H$_2$-He and Rayleigh scattering by H$_2$ and He.  

The simulation was run for 1000 Earth days, and the final output is used for the analyses here. Extended details of the GCM setup are presented in Table \ref{tab: GCMParams}.

\begin{table}
 \footnotesize
 \caption{GCM Parameters}
 \label{tab: GCMParams}
 \begin{tabular}{lc}
 \hline
 \hline
    & Value \\
 \hline
 {\em Grid and Domain} \\
 Longitude Cells & 144  \\
 Latitude Cells & 90  \\
 Vertical Layers & 86  \\
 Domain Height &  $1.01\times 10^7$ m \\
 Domain Inner Radius  & $7.86\times 10^7$ m  \\
 Hydrodynamic Time step & 30 s \\
 \\
 {\em Radiative Transfer} \\
 Wavelength Bins & 32 \\
 Wavelength Minimum  & 0.2 $\mathrm{\mu m}$\\
 Wavelength Maximum  & 322 $\mathrm{\mu m}$\\
 Radiative Time step & 150 s\\
 \\
 {\em Damping and Diffusion} \\
 Damping Profile & Horizontally Uniform  \\
 Damping Coefficient & 0.15   \\
 Damping Depth ($\eta_s$) & 0.8   \\
 \\
 {\em Planet}\\
 Intrinsic Temperature & 100 K \\
 Initial Inner Boundary Pressure & 200 bar\\
 Semi-major axis $a$  & $4.53\times 10^{-2}$ AU \\
 Stellar Constant at 1 AU &  1272 $\mathrm{W\, m^{-2}}$ \\
 Specific gas constant R  &  3164.7 $\mathrm{J\,kg^{-1}K^{-1}}$ \\
 Specific heat capacity $c_\mathrm{P}$ & 11476.7 $\mathrm{J\,kg^{-1}K^{-1}}$\\
 g at inner boundary  & 10.04 $\mathrm{m\,s^{-2}}$\\
 \hline
 \end{tabular}
\end{table}

\section{Retrieval Summaries and Example Corner Plots}
\label{app: example corner}

\begin{figure*}
    \centering
    \includegraphics[width=0.95\linewidth]{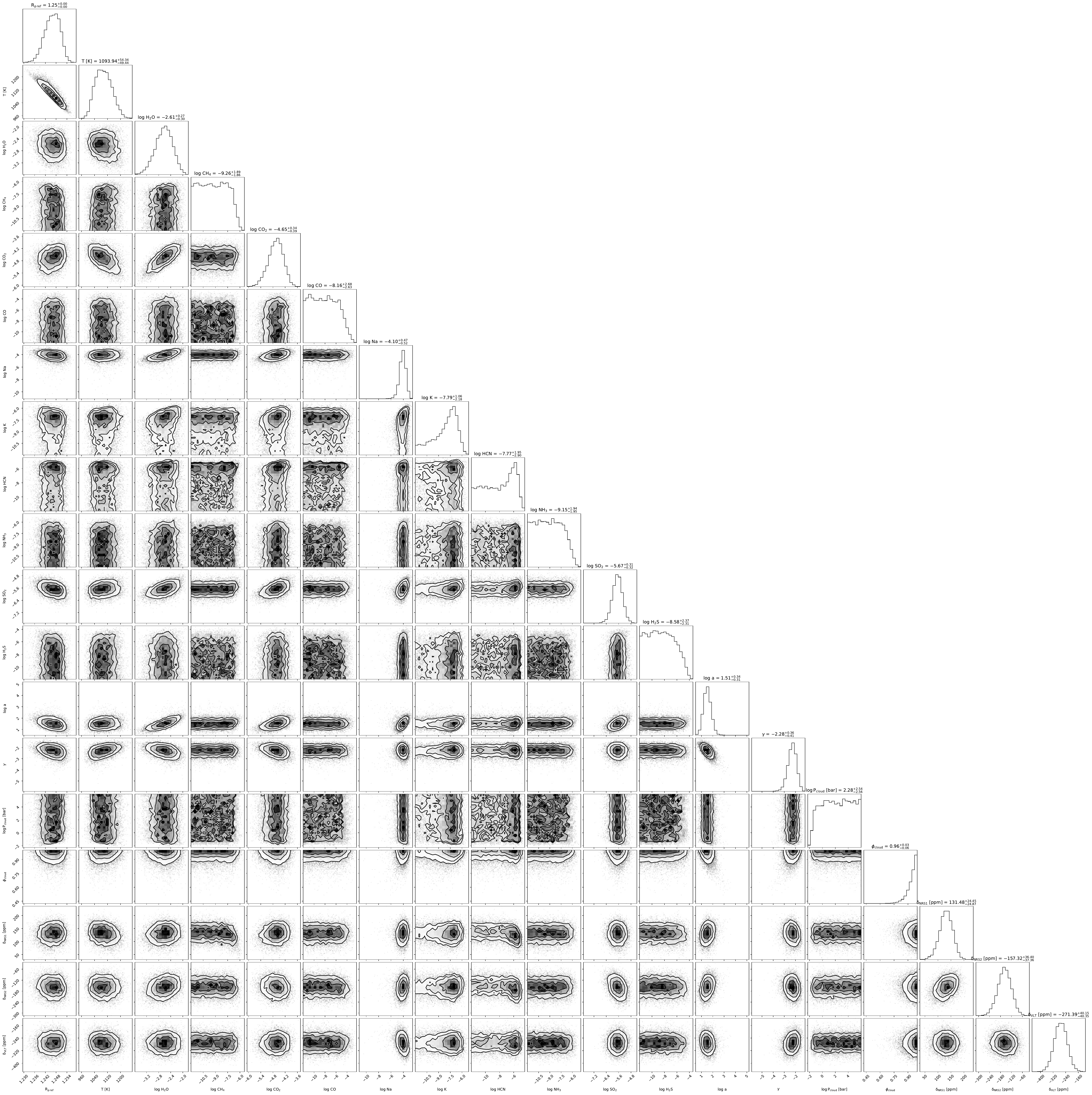}
    \caption{Corner plot from the \texttt{POSEIDON} retrieval on the combined VLT + \texttt{exoTEDRF} $R=300$ JWST spectrum.}
    \label{fig:poseidon_corner}
\end{figure*}

Table~\ref{tab: Opacities} has citations for all opacity source used in our atmosphere models. Figure~\ref{fig:poseidon_corner} shows the corner plot from the \texttt{POSEIDON} retrieval on the combined $R=300$ \texttt{exoTEDRF} JWST + VLT spectrum. Corner plots from all other retrievals are available on Zenodo. 

\begin{deluxetable*}{c||ccccc}
 \tabletypesize{\scriptsize}
 \tablecaption{Opacity Source and References }  
 \label{tab: Opacities}
 \tablehead{
  Parameter & \texttt{POSEIDON} & \texttt{NemesisPy} & \texttt{Pyrat Bay} & \texttt{Aurora} & \texttt{ScCHIMERA}
           }
    \startdata
     H$_2$O & \citet{polyansky_exomol_2018} & `` " & \citet{rothman_hitemp_2010} & \citet{polyansky_exomol_2018} & `` " \\
     CO$_2$ & \citet{yurckenko_exomol_2020} & `` " & \citet{rothman_hitemp_2010} & `` " & \citet{Freedman2014} \\
     CO & \citet{li_rovibrational_2015} & `` " & `` " & \citet{rothman_hitemp_2010} & `` " \\
     CH$_4$ & \citet{yurchenko_exomol_2024} & `` " & \citet{Hargreaves_2020} & \citet{yurchenko_exomol_2024} & \citet{rothman_hitemp_2010} \\
     SO$_2$ & \citet{underwood_exomol_2016} & `` " & `` " & `` " & `` " \\
     H$_2$S & \citet{azzam_exomol_2016} & `` " & - & - & `` " \\
     HCN & \citet{barber_exomol_2014} & `` " & \citet{harriss_20086} & \citet{barber_exomol_2014} & `` " \\
     NH$_3$ & \citet{coles_exomol_2019} & `` " & \citet{yurchenko_variationally_2011} & `` " & \citet{coles_exomol_2019} \\
     Na & \citet{ryabchikova_major_2015} & \citet{Allard2019} & `` " & `` " & `` "\\
     K & \citet{ryabchikova_major_2015} & \citet{Allard2016} & `` " & `` " & `` "\\
     C$_2$H$_2$ & - & - & - & - & \citet{Chubb2020} \\
     H$_2$-H$_2$ & \citet{Chubb2021} & \citet{Borysow2001}, & `` " & \citet{richard_new_2012} & \citet{Freedman2008} \\
      & & \citet{Borysow2002} & & & \\
     H$_2$-He & \citet{Chubb2021} & \citet{Borysow1989}, & `` " & \citet{richard_new_2012} & \citet{Freedman2008} \\
      & & \citet{Borysow1989b} & & & \\[2pt]
    \enddata
    \tablecomments{ - indicates that a parameter was not included in a retrieval.
                    `` " denotes repeat of the previous column.}
\end{deluxetable*}

\bibliography{main}{}
\bibliographystyle{aasjournal}

\end{document}